\documentclass[journal,12pt,onecolumn,draftclsnofoot]{IEEEtran}
\usepackage{graphicx}
\usepackage[center]{caption}
\usepackage{subcaption}
\usepackage{cite}
\usepackage{dsfont}
\usepackage{color}
\usepackage{amsmath,amssymb,amsthm}
\usepackage{lettrine}
\newcommand\norm[1]{\left\lVert#1\right\rVert}
\usepackage{algorithm,algorithmic,multicol}

\newtheoremstyle{mystyle}
{}
{}
{\itshape}
{}
{\bfseries}
{.}
{ }
{}
\theoremstyle{mystyle}

\begin{document}
\title{Joint Bit Allocation and Hybrid Beamforming Optimization for Energy Efficient Millimeter Wave MIMO Systems}

\author{Aryan Kaushik, Evangelos Vlachos, \IEEEmembership{Member,~IEEE,} \\Christos Tsinos, \IEEEmembership{Member,~IEEE,} John Thompson, \IEEEmembership{Fellow,~IEEE,} \\ Symeon Chatzinotas, \IEEEmembership{Senior Member,~IEEE}

\thanks{Aryan Kaushik is with the Department of Electronic and Electrical Engineering, University College London, U.K. (e-mail: a.kaushik@ucl.ac.uk). Evangelos Vlachos is with the Industrial Systems Institute, Athena Research Centre, Patras, Greece. (e-mail: evlachos@isi.gr). John Thompson is with the Institute for Digital Communications, The University of Edinburgh, U.K. (e-mail: j.s.thompson@ed.ac.uk). Christos Tsinos and Symeon Chatzinotas are with the Interdisciplinary Centre for Security, Reliability and Trust, University of Luxembourg, Luxembourg (e-mail: \{christos.tsinos, symeon.chatzinotas\}@uni.lu).}
\thanks{The Engineering and Physical Sciences Research Council under Grant EP/P000703/1 and ECLECTIC project under FNR CORE Framework supported this work partly.
A part of the content of this paper is presented at 2019 IEEE GLOBECOM\cite{aryanGC2019}.}
}
\maketitle
\vspace{-15mm}

\begin{abstract}
In this paper, we aim to design highly energy efficient end-to-end communication for millimeter wave multiple-input multiple-output systems. This is done by jointly optimizing the digital-to-analog converter (DAC)/analog-to-digital converter (ADC) bit resolutions and hybrid beamforming matrices. The novel decomposition of the hybrid precoder and the hybrid combiner to three parts is introduced at the transmitter (TX) and the receiver (RX), respectively, representing the analog precoder/combiner matrix, the DAC/ADC bit resolution matrix and the baseband precoder/combiner matrix. The unknown matrices are computed as a solution to the matrix factorization problem where the optimal fully digital precoder or combiner is approximated by the product of these matrices. A novel and efficient solution based on the alternating direction method of multipliers is proposed to solve these problems at both the TX and the RX. The simulation results show that the proposed solution, where the DAC/ADC bit allocation is dynamic during operation, achieves higher energy efficiency when compared with existing benchmark techniques that use fixed DAC/ADC bit resolutions.
\end{abstract}


\IEEEpeerreviewmaketitle

\section{Introduction}
\lettrine{\textbf{M}}{illimeter wave} (mmWave) spectrum is an attractive alternative to the densely occupied microwave spectrum range of 300 MHz to 6 GHz for next generation wireless communication systems. The advantages of using a mmWave frequency band are increased capacity, 
lower latency, high mobility and reliability, and lower infrastructure costs \cite{andJSAC2014, rappaportACCESS2013, boccardiCM2014}. The higher path loss associated with mmWave spectrum can be compensated by using large scale antenna arrays leading to a multiple-input multiple-output (MIMO) system. Implementing fully digital beamforming in mmWave MIMO systems provides high throughput but has high complexity and low energy efficiency (EE). A simpler alternative is a fully analog beamforming approach which was discussed in \cite{ayachSPAWC2012} but cannot implement multi-stream spatial communication due to the use of a single radio frequency (RF) chain. 

Analog/digital (A/D) hybrid beamforming MIMO architectures implement both digital and analog units to overcome these issues. The hardware complexity and power consumption is reduced through using fewer RF chains and it can support multi-stream communication with high spectral efficiency (SE) \cite{ayachTWC2014, aryanIET2016, hanCM2015, bogaleTWC2016, payamiTWC2016, payamiTVT2018, liCL2017, TsinosTSP2018, TsinosASILOMAR2016, TsinosTCCN2019}. Such systems can be also optimized to achieve high EE gains \cite{ranziSAC2016, TsinosSAC2017, aryanTGCN2019, aryanComNet2020}. An alternative solution to reduce the power consumption and hardware complexity is by decreasing the bit resolution \cite{heathSSP2016} of the digital-to-analog converters (DACs) and the analog-to-digital converters (ADCs). 
Given the distinct system and channel model characteristics at mmWave compared to microwave, the EE and SE performance needs to be analyzed for the A/D hybrid beamforming architecture with low resolution sampling. 


\subsection{Literature Review}
To observe the effect of ADC resolution and bandwidth on rate, an additive quantization noise model (AQNM) is considered in \cite{orhanITA2015} for a mmWave MIMO system under a RX power constraint. Reference \cite{fanCL2015} uses AQNM and shows the significance of low resolution ADCs on decreasing the rate. Recent work on A/D hybrid MIMO systems with low resolution sampling dynamically adjusts the ADC resolution \cite{choiTSP2017}. Most of the literature such as in \cite{orhanITA2015,fanCL2015,choiTSP2017,jmoITG2016,zhangSAC2017,aryanEUSIPCO2018,tczhangTWC2016} imposes low resolution only at the RX side, and mostly assumed a fully digital or hybrid TX with high resolution DACs. However, there is a need to conduct research on optimizing the bit resolution problem for the TX side as well.

Furthermore, the existing literature mostly develops systems based on high resolution ADCs with a small number of RF chains or low resolution ADCs with a large number of RF chains. Either way, only fixed resolution DACs/ADCs are taken into account. References \cite{ranziSAC2016, TsinosSAC2017} consider EE optimization problems for A/D hybrid transceivers but with fixed and high resolution at the DACs/ADCs. The power model in \cite{ranziSAC2016} takes into account the power consumed at every RF chain and a constant power term for site-cooling, baseband processing and synchronization at the TX and \cite{TsinosSAC2017} considers the RF hardware losses and some computational power expenditure. 

Some approaches have been applied in A/D hybrid mmWave MIMO systems for EE maximization and low complexity with both full and low resolution sampling cases \cite{aryanTGCN2019, aryanComNet2020, aryanICC2019}. Reference \cite{aryanTGCN2019} proposes an energy efficient A/D hybrid beamforming framework with a novel architecture for a mmWave MIMO system. The number of active RF chains are optimized dynamically by fractional programming to maximize EE performance but the DAC/ADC bit resolutions are fixed. Reference \cite{aryanICC2019} proposes a novel EE maximization technique that selects the best subset of the active RF chains and DAC resolution which can also be extended to low resolution ADCs at the RX. Reference \cite{jmoITG2016} suggests implementing fixed and low resolution ADCs with a small number of RF chains. Reference \cite{zhangSAC2017} works on the idea of a mixed-ADC architecture where a better energy-rate trade off is achieved by combining low and high resolution ADCs, but still with a fixed resolution for each ADC and without considering A/D hybrid beamforming. An A/D hybrid beamforming system with fixed and low resolution ADCs has been analyzed for channel estimation in \cite{aryanEUSIPCO2018}. 

One can implement varying resolution ADCs at the RX \cite{tczhangTWC2016} which may provide a better solution than the RX with fixed and low resolution ADCs. Similarly, exploring low resolution DACs at the TX can also help reduce the power consumption. Thus, research that is focused on ADCs at the RX can also be applied to the TX DACs considering the TX specific system model parameters.  Similar to using different ADC resolutions at the RX \cite{tczhangTWC2016}, which could provide a better solution than fixed low resolution ADCs, one can design a variable DAC resolution TX. Extra care is needed when deciding the number of bits used as the total DAC/ADC power consumption can be dominated by only a few high resolution DACs/ADCs. From \cite{singhTWC2009}, we notice that a good trade off between the power consumption and the performance may be to consider the range of 1-8 bits for I- and Q-channels, 
where 8-bit represents the full-bit resolution DACs/ADCs.

Reference \cite{mezghaniECS2009} uses low resolution DACs for a single user MIMO system while \cite{jacobssonTC2017} employs low resolution DACs at the base station for a narrowband multi-user MIMO system. Reference \cite{TsinosSPAWC2018} also discusses fixed and low resolution DACs architecture for multi-user MIMO systems. Reference \cite{ribeiroTSP2018} considers a single user MIMO system with quantized hybrid precoding including the RF quantized noise term beside the additive white Gaussian noise (AWGN) while evaluating EE and SE performance. The existing literature still does not consider adjusting the resolution associated with DACs/ADCs dynamically. It is possible to consider both the TX and the RX simultaneously where we can design an optimization problem to find the optimal number of quantized bits to achieve high EE performance. When designing for high EE, the complexity of the solution also needs to be taken into account while providing improvements over the existing literature. 

\subsection{Contributions}
This paper designs an optimal EE solution for a mmWave A/D hybrid MIMO system by introducing a novel TX decomposition of the A/D hybrid precoder to three parts representing the analog precoder matrix, the DAC bit resolution matrix and the digital precoder matrix, respectively. A similar decomposition at the RX represents the analog combiner matrix, the ADC bit resolution matrix and the digital combiner matrix. 
Our aim is to minimize the distance between the decomposition, which is expressed as the product of three matrices, and the corresponding fully digital precoder or combiner matrix. The joint problem is decomposed into a series of sub-problems which are solved using 
the alternating direction method of multipliers (ADMM). 
We implement an exhaustive search approach \cite{ranziSAC2016} to evaluate the upper bound for EE maximization. 

In \cite{aryanGC2019}, we addressed bit allocation and hybrid combining at the RX only, where we jointly optimized the number of ADC bits and hybrid combiner matrices for EE maximization. A novel decomposition of the hybrid combiner to three parts was introduced: the analog combiner matrix, the bit resolution matrix and the baseband combiner matrix, and these matrices were computed using the ADMM approach in order to solve the matrix factorization problem. In addition to \cite{aryanGC2019}, the main contributions of this paper can be listed as follows:
\begin{itemize}
    \item This paper designs an optimal EE solution for a mmWave A/D hybrid beamforming MIMO system by introducing a novel matrix decomposition applied to the hybrid beamforming matrices at both the TX and the RX. This decomposition defines three matrices, which are the analog beamforming matrix, the bit resolution matrix and the baseband beamforming matrix at both the TX and the RX. These matrices are obtained by the solution of an EE maximization problem and the DAC/ADC bit resolution is adjusted dynamically unlike the fixed bit resolution \cite{fanCL2015, jmoITG2016}, considered in the existing literature.
    \item The joint TX-RX problem is a difficult problem to solve due to non-convex constraints and the non-convex cost function. 
    First, we decouple it into two sub-problems dealing with the TX and the RX separately. The corresponding problems at the TX and the RX are solved by a novel algorithmic solution based on ADMM to obtain the unknown precoder/combiner and DAC/ADC bit resolution matrices.
    \item Thus, this work jointly optimizes the hybrid beamforming and DAC/ADC bit resolution matrices, unlike the existing approaches that optimize either DAC/ADC bit resolution \cite{aryanICC2019} or hybrid beamforming matrices \cite{aryanTGCN2019, aryanComNet2020}. Moreover, the proposed design has high flexibility, given that the analog precoder/combiner is codebook-free, thus there is no restriction on the angular vectors and different bit resolutions can be assigned to each DAC/ADC.
\end{itemize}

The performance of the proposed technique is investigated through extensive simulation results, achieving increased EE compared to the baseline techniques with fixed DAC/ADC bit resolutions and number of RF chains, and an exhaustive search based approach which is an upper bound for EE maximization.

\subsection{Notation and Organization} 
$\mathbf{A}$, $\mathbf{a}$ and $\textit{a}$ stand for a matrix, a vector, and a scalar, respectively. The trace, transpose  and complex conjugate transpose of $\mathbf{A}$ are denoted as $\textrm{tr}(\mathbf{A})$, $\mathbf{A}^{T}$ and $\mathbf{A}^{H}$, respectively; $\Vert \mathbf{A} \Vert_{F}$ represents the Frobenius norm of $\mathbf{A}$; $|a|$ represents the determinant of $a$; 
$\mathbf{I}_{N}$ represents $N \times N$ identity matrix; $\mathcal{C}\mathcal{N} (\mathbf{a}; \mathbf{A})$ denotes a complex Gaussian vector having mean $\mathbf{a}$ and covariance matrix $\mathbf{A}$; $\mathbb{C}$, $\mathbb{R}$ and $\mathbb{R}^+$ denote the sets of complex numbers, real numbers and positive real numbers, respectively; $\textbf{X} \in \mathbb{C}^{A \times B}$ and $\textbf{X} \in \mathbb{R}^{A \times B}$ denote $A \times B$ size $\mathbf{X}$ matrix with complex and real entries, respectively; $[\mathbf{A}]_k$ denotes the $k$-th column of matrix $\mathbf{A}$ while $[\mathbf{A}]_{kl}$ the matrix entry at the $k$-th row and $l$-th column; the indicator function $\mathds{1}_{\mathcal{S}}\left\{\mathbf{A}\right\}$ of a set $\mathcal{S}$ that acts over a matrix $\mathbf{A}$  is defined as $0 \hspace{1mm} \forall \hspace{1mm} \mathbf{A} \in \mathcal{S}$ and $\infty \hspace{1mm} \forall \hspace{1mm} \mathbf{A} \notin \mathcal{S}$.

Section II presents the channel and system models where the channel model is based on a mmWave channel setup and the system model defines the low resolution quantization at both the TX and the RX. Sections III and IV present the problem formulation for the proposed technique at the TX and the RX, respectively, and the solution to obtain an energy efficient system. Section V verifies the proposed technique through simulation results and Section VI concludes the paper.

\begin{figure*}[t]
\centering
\begin{subfigure}{\linewidth}
\centering
   \includegraphics[width=0.85\textwidth, trim=170 120 55 150,clip]{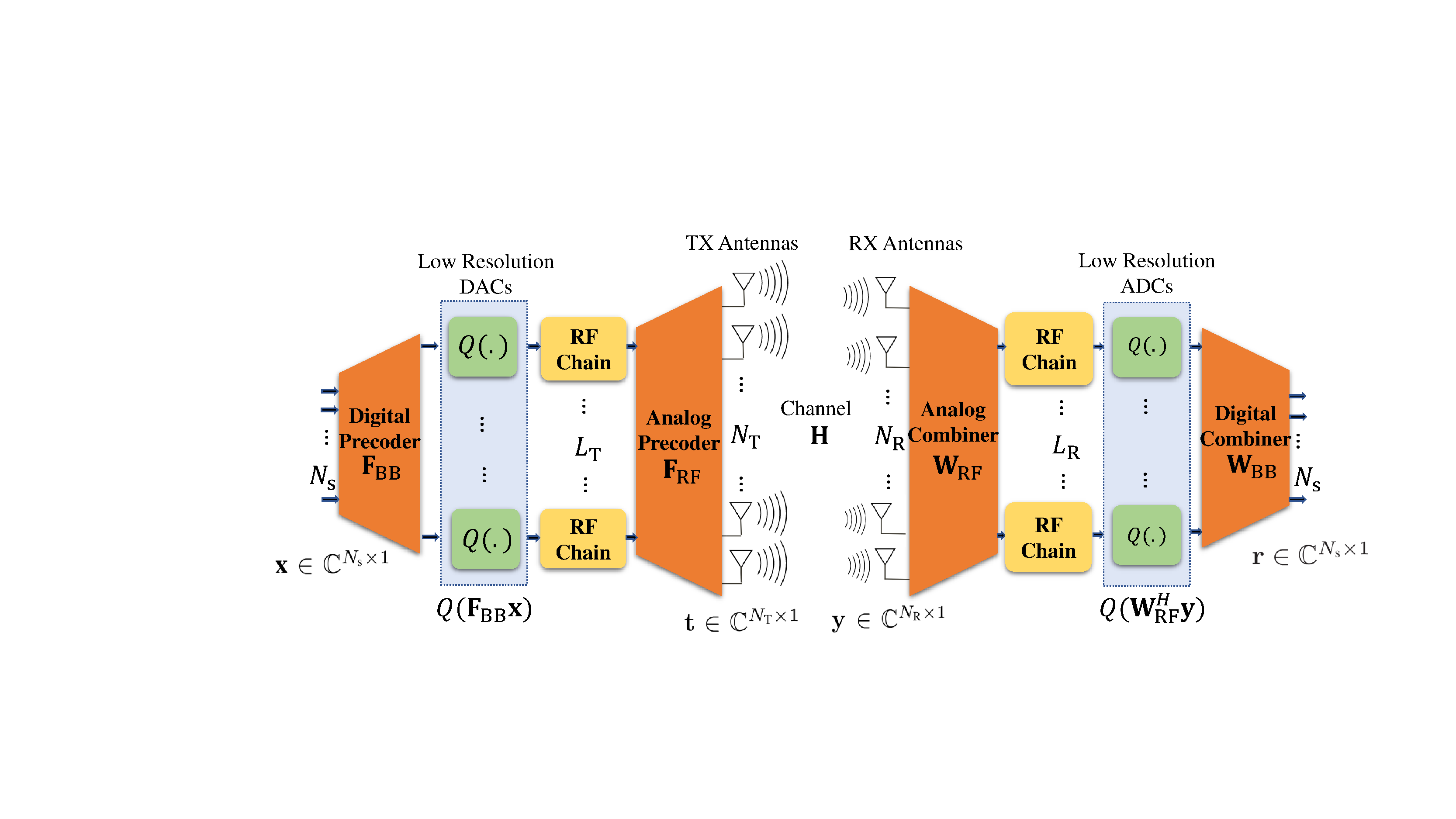}
		\caption{A mmWave A/D hybrid MIMO system with varying DAC/ADC bit resolutions at the TX/RX.}
\end{subfigure}
\begin{subfigure}{\linewidth}
\centering
    \includegraphics[width=0.8\textwidth, trim=100 140 80 200,clip]{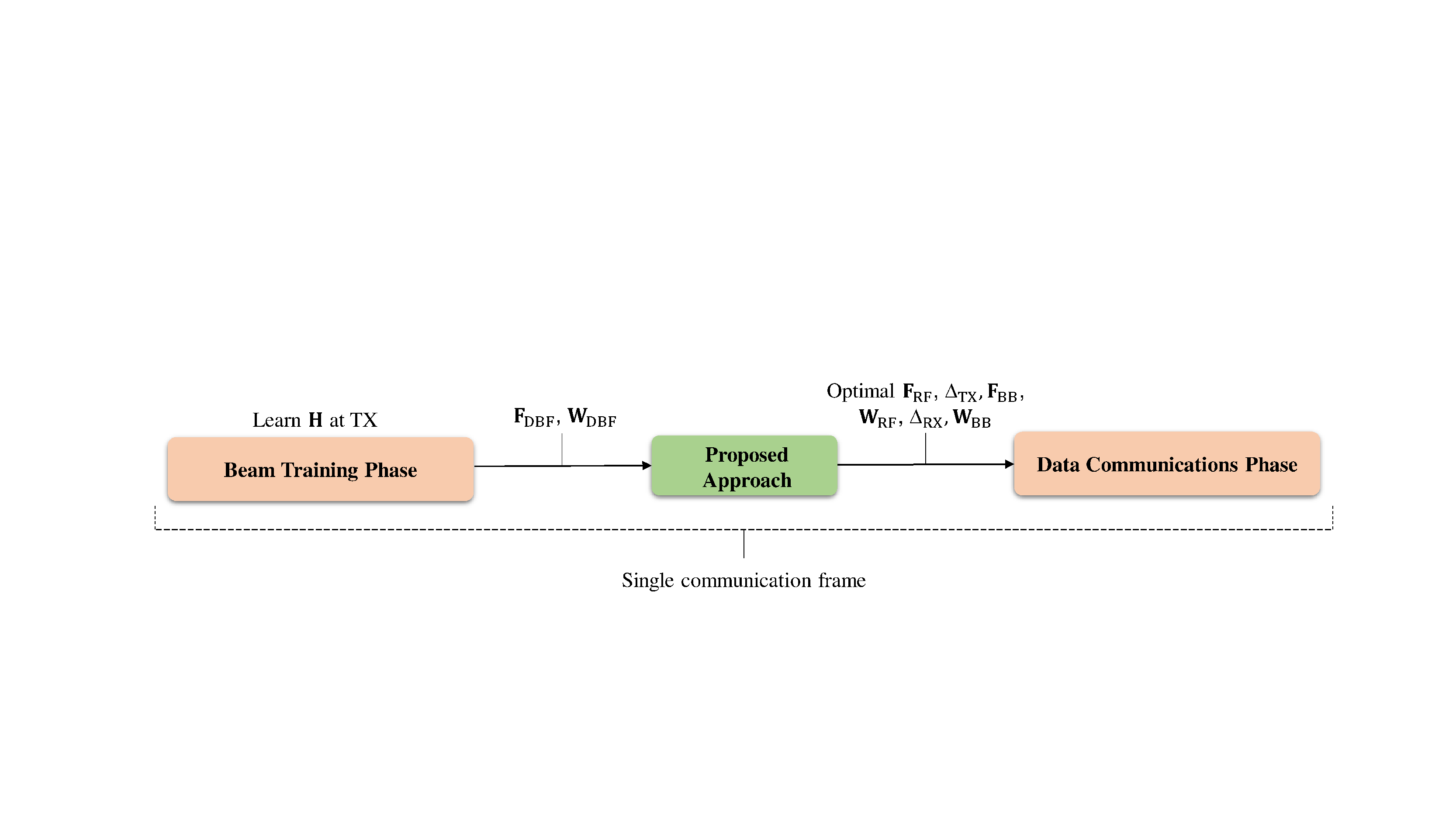}
		\caption{Block diagram of the beam tracking phase and the data communications phase.}
\end{subfigure}
\caption{System model for mmWave hybrid MIMO with varying DAC/ADC bit resolution.}
\vspace{-3mm}
\end{figure*}

\section{MmWave A/D Hybrid MIMO System}
\subsection{MmWave Channel Model}
MmWave channels can be modeled by a narrowband clustered channel model due to different channel settings such as the number of multipaths, amplitudes, etc., with $N_{\textrm{cl}}$ clusters and $N_{\textrm{ray}}$ propagation paths in each cluster \cite{ayachTWC2014}. Considering a single user mmWave system with $N_{\textrm{T}}$ antennas at the TX, transmitting $N_\textrm{s}$ data streams to $N_\textrm{R}$ antennas at the RX, the mmWave channel matrix can be written as follows:
\begin{equation}\label{eq:channel_model}
\mathbf{H} = \sqrt{\frac{N_\textrm{T}N_\textrm{R}}{N_\textrm{cl} N_\textrm{ray}}} \sum_{i=1}^{N_{\textrm{cl}}} \sum_{l=1}^{N_{\textrm{ray}}} \alpha_{il} \mathbf{a}_{\textrm{R}}(\phi_{il}^{r}) \mathbf{a}_{\textrm{T}}(\phi_{il}^{t})^H,
\end{equation}
where $\alpha_{il} \in \mathcal{C}\mathcal{N}(0,\sigma_{\alpha,i}^2)$ is the gain term with $\sigma_{\alpha,i}^2$ being the average power of the $i^{th}$ cluster. Furthermore, $\mathbf{a}_{\textrm{T}}(\phi_{il}^{t})$ and $\mathbf{a}_{\textrm{R}}(\phi_{il}^{r})$ represent the normalized transmit and receive array response vectors \cite{ayachTWC2014}, where $\phi_{il}^{t}$ and $\phi_{il}^{r}$ denote the azimuth angles of departure and arrival, respectively. We use uniform linear array (ULA) antennas for simplicity and model the antenna elements at the RX as ideal sectored elements \cite{singh}. We assume that the channel state information (CSI) is known at both the TX and the RX.

\subsection{A/D Hybrid MIMO System Model}
Based on the A/D hybrid beamforming scheme in the large scale mmWave MIMO communication systems, the number of TX RF chains $L_\textrm{T}$ follows the limitation $N_\textrm{s} \leq L_\textrm{T} \leq N_\textrm{T}$ and similarly for $L_\textrm{R}$ RF chains at the RX, $N_\textrm{s} \leq L_\textrm{R} \leq N_\textrm{R}$ \cite{ayachTWC2014, aryanIET2016}. As shown in Fig. 1 (a), the matrices $\mathbf{F}_{\textrm{RF}} \in \mathbb{C}^{N_\textrm{T}\times L_\textrm{T}}$ and $\mathbf{F}_{\textrm{BB}} \in \mathbb{C}^{L_\textrm{T}\times N_\textrm{s}}$ denote the analog precoder and baseband precoder matrices, respectively. Similarly, the matrices $\mathbf{W}_{\textrm{RF}} \in \mathbb{C}^{N_\textrm{R}\times L_\textrm{R}}$ and $\mathbf{W}_{\textrm{BB}} \in \mathbb{C}^{L_\textrm{R}\times N_\textrm{s}}$ denote the analog combiner and baseband combiner matrices, respectively. The analog precoder and combiner matrices, $\mathbf{F}_{\textrm{RF}}$ and $\mathbf{W}_{\textrm{RF}}$, are based on phase shifters, i.e., the elements that have unit modulus and continuous phase. Thus, $\mathbf{F}_\textrm{RF} \in \mathcal{F}^{N_\textrm{T} \times L_\textrm{T}}$ and $\mathbf{W}_\textrm{RF} \in \mathcal{W}^{N_\textrm{R} \times L_\textrm{R}}$ where the set $\mathcal{F}$ and $\mathcal{W}$ represent the set of possible phase shifts in $\mathbf{F}_\textrm{RF}$ and $\mathbf{W}_\textrm{RF}$, respectively. The sets $\mathcal{F}$ and $\mathcal{W}$ for variables $f$ and $w$, respectively, are defined as $\mathcal{F} = \left\{f \in \mathbb{C} \ | \ |f| = 1\right\}$ and $\mathcal{W} = \left\{w \in \mathbb{C} \ | \ |w| = 1\right\}$.

Note that, we optimize the DAC and ADC resolution and the precoder and combiner matrices at the TX and the RX on a frame-by-frame basis. As shown in Fig. 1 (b), we consider two stages in the system model: i) the beam training phase, and ii) the data communications phase. In stage i), firstly, the channel $\mathbf{H}$ is computed which provides us the optimal beamforming matrices, i.e., $\mathbf{F}_\textrm{DBF}$ at the TX and $\mathbf{W}_\textrm{DBF}$ at the RX. 
In stage ii), the optimal precoding and DAC bit resolution matrices $\mathbf{F}_\textrm{RF}$, $\mathbf{F}_\textrm{BB}$ and $\mathbf{\Delta}_\textrm{TX}$ at the TX, respectively, and the optimal combining and ADC bit resolution matrices $\mathbf{W}_\textrm{RF}$, $\mathbf{W}_\textrm{BB}$ and $\mathbf{\Delta}_\textrm{RX}$ at the RX are obtained. These two phases consist of one communication frame where the frame duration is smaller than the channel coherence time. Furthermore, if we assume that the TX/RX is active for stage i) a small proportion of time, for example, $<10\%$, then the overall transmit energy consumption is dominated by stage ii). 

We consider the linear AQNM to represent the distortion of quantization \cite{orhanITA2015}. Given that $Q(\cdot)$ denotes a uniform scalar quantizer then for the scalar complex input $x \in \mathbb{C}$ that is applied to both the real and imaginary parts, we have, $Q(x) \approx \delta x + \epsilon$,
where $\delta = \sqrt{1-\frac{\pi \sqrt{3}}{2}2^{-2 b}} \in [m,M]$ is the multiplicative distortion parameter for a bit resolution equal to $b$ \cite{mezghaniISIT2012}, where $m$ and $M$ denote the minimum and maximum value of the range. The resolution parameter $b$ is denoted as $b_i^t \, \forall \, i=1,\ldots, L_\textrm{T}$ and $b_i^r \, \forall \, i=1,\ldots, L_\textrm{R}$ at the TX and the RX, respectively. Note that the introduced error in the above linear approximation 
decreases for larger resolutions. However, our proposed solution focuses on EE maximization and this linear approximation does not impact the performance significantly as observed from the simulation results in Section V. The parameter $\epsilon$ is the additive quantization noise with $\epsilon \sim \mathcal{CN}(0, \sigma_{\epsilon}^2)$, where  $\sigma_{\epsilon} = \sqrt{1-\frac{\pi \sqrt{3}}{2}2^{-2 b}} \sqrt{\frac{\pi \sqrt{3}}{2}2^{-2 b}}$. The matrices $\boldsymbol{\Delta}_\textrm{TX}$ and $\boldsymbol{\Delta}_\textrm{RX}$ represent diagonal matrices with values depending on the bit resolution of each DAC and ADC, respectively. Specifically, each diagonal entry of $\mathbf{\Delta}_\textrm{TX}$ is given by:
\begin{equation}\label{eq:tx_delta_distorsion_rf}
[\boldsymbol{\Delta}_\textrm{TX}]_{ii} = \sqrt{1-\frac{\pi \sqrt{3}}{2}2^{-2 b_i^t}} \in [m,M] \hspace{1mm} \forall \hspace{1mm} i=1,\ldots, L_\textrm{T},
\end{equation}
and each diagonal entry of $\boldsymbol{\Delta}_\textrm{RX}$ is given by:
\begin{equation}\label{eq:rx_delta_distorsion_rf}
[\boldsymbol{\Delta}_\textrm{RX}]_{ii} = \sqrt{1-\frac{\pi \sqrt{3}}{2}2^{-2 b_i^r}} \in [m,M] \hspace{1mm} \forall \hspace{1mm} i=1,\ldots, L_\textrm{R},
\end{equation}
where, for simplicity, we assume that the range $[m,M]$ is the same for each of the DACs/ADCs. The additive quantization noise for the DACs and ADCs are written as complex Gaussian vectors $\boldsymbol{\epsilon}_\textrm{TX} \in \mathcal{C}\mathcal{N}(\mathbf{0}, \mathbf{C}_{\epsilon\textrm{T}})$ and $\boldsymbol{\epsilon_\textrm{RX}} \in \mathcal{C}\mathcal{N}(\mathbf{0}, \mathbf{C}_{\epsilon\textrm{R}})$ \cite{aryanICC2019} 
where $\mathbf{C}_{\epsilon\textrm{T}}$ and $\mathbf{C}_{\epsilon\textrm{R}}$ are the diagonal covariance matrices for DACs and ADCs, respectively. The covariance matrix entries are as follows: 
\begin{equation}
    [\mathbf{C}_{\epsilon\textrm{T}}]_{ii} \!=\! \left(1\!-\!\frac{\pi \sqrt{3}}{2}2^{-2 b_i^t}\right)\!\left(\frac{\pi \sqrt{3}}{2}2^{-2 b_i^t}\right) \! \forall\! \hspace{1mm} i\!=\!1,.., L_\textrm{T},
\end{equation}
and
\begin{equation}
    [\mathbf{C}_{\epsilon\textrm{R}}]_{ii} \!=\! \left(1\!-\!\frac{\pi \sqrt{3}}{2}2^{-2 b_i^r}\right)\!\left(\frac{\pi \sqrt{3}}{2}2^{-2 b_i^r}\right) \! \forall\! \hspace{1mm} i\!=\!1,.., L_\textrm{R}.
\end{equation}
Note that while optimizing the EE of the TX side, it is considered that the RX parameters, which includes the analog combiner matrix, the ADC bit resolution matrix and the baseband combiner matrix is known to the TX and vice-versa. 

Let us consider $\mathbf{x} \in \mathbb{C}^{N_\textrm{s} \times 1}$ as the normalized data vector, then based on the AQNM, the vector containing the complex output of all the DACs can be expressed as follows:
\begin{equation}\label{eq:tx_quantization_mimo_model}
Q(\mathbf{F}_{\textrm{BB}} \mathbf{x}) \approx \boldsymbol{\Delta}_\textrm{TX} \mathbf{F}_{\textrm{BB}} \mathbf{x} + \boldsymbol{\epsilon}_\textrm{TX} \in \mathbb{C}^{L_{\textrm{T}} \times 1}, 
\end{equation}
This leads us to the following linear approximation for the transmitted signal $\mathbf{t} \in \mathbb{C}^{N_\textrm{T} \times 1} $, as seen at the output of the A/D hybrid TX in Fig. 1 (a):
\begin{equation}\label{eq:transmitted_signal}
\mathbf{t} = \mathbf{F}_{\textrm{RF}} \boldsymbol{\Delta}_\textrm{TX} \mathbf{F}_{\textrm{BB}} \mathbf{x} + \mathbf{F}_{\textrm{RF}}  \boldsymbol{\epsilon}_\textrm{TX}.
\end{equation}

After the effect of the wireless mmWave channel $\mathbf{H}$ and the Gaussian noise $\mathbf{n}$ with independent and identically distributed entries and complex Gaussian distribution, i.e., $\mathbf{n} \sim \mathcal{C}\mathcal{N}(\mathbf{0},\sigma_\textrm{n}^2 \mathbf{I}_{N_{\textrm{R}}})$, the received signal $\mathbf{y} \in \mathbb{C}^{N_\textrm{R}\times1}$ is expressed as follows:
\begin{align}\label{eq:received_signal}
\mathbf{y} = & \mathbf{H} \mathbf{t} + \mathbf{n} = \mathbf{H}\mathbf{F}_{\textrm{RF}} \boldsymbol{\Delta}_\textrm{TX} \mathbf{F}_{\textrm{BB}} \mathbf{x} + \mathbf{H} \mathbf{F}_{\textrm{RF}}  \boldsymbol{\epsilon}_\textrm{TX} + \mathbf{n}.
\end{align}
When the analog combiner matrix $\mathbf{W}_\textrm{RF}$ and ADC quantization based on AQNM are applied to the received signal $\mathbf{y}$, we obtain the following:
\begin{equation}\label{eq:rx_quantization_mimo_model}
Q(\mathbf{W}_\textrm{RF}^H\mathbf{y} ) \approx \boldsymbol{\Delta}_\textrm{RX}^H \mathbf{W}_\textrm{RF}^H \mathbf{y} + \boldsymbol{\epsilon}_\textrm{RX} \in \mathbb{C}^{L_{\textrm{R}} \times 1}.
\end{equation}

After the application of the baseband combiner matrix $\mathbf{W}_\textrm{BB}$, the output signal $\mathbf{r} \in \mathbb{C}^{N_\textrm{s} \times 1}$ at the RX, as shown in Fig. 1 (a), can be expressed as follows:
\begin{align}\label{eq:receiver_output_initial}
\mathbf{r} =  
\mathbf{W}_\textrm{BB}^H \mathbf{\Delta}_\textrm{RX}^H \mathbf{W}_\textrm{RF}^H \mathbf{y} + \mathbf{W}_\textrm{BB}^H \boldsymbol{\epsilon}_\textrm{RX}. 
\end{align}
Considering the A/D hybrid precoder matrix $\mathbf{F}\! =\! \mathbf{F}_\textrm{RF}\boldsymbol{\Delta}_\textrm{TX}\mathbf{F}_\textrm{BB} \!\in\! \mathbb{C}^{N_\textrm{T} \times N_\textrm{s}}$ and the A/D hybrid combiner matrix $\mathbf{W} \!=\! \mathbf{W}_\textrm{RF}\boldsymbol{\Delta}_\textrm{RX}\mathbf{W}_\textrm{BB} \!\in\! \mathbb{C}^{N_\textrm{R} \times N_\textrm{s}}$, we can express the RX output signal $\mathbf{r}$ in \eqref{eq:receiver_output_initial} as follows:
\begin{equation}\label{eq:tx_system_model}
\mathbf{r} = \mathbf{W}^H \mathbf{H} \mathbf{F} \mathbf{x} + \underbrace{\mathbf{W}^H \mathbf{H} \mathbf{F}_{\textrm{RF}}  \boldsymbol{\epsilon}_\textrm{TX}  +\mathbf{W}_\textrm{BB}^H \boldsymbol{\epsilon}_\textrm{RX} + \mathbf{W}^H\mathbf{n}}_{\boldsymbol{\eta}}, 
\end{equation}
where $\boldsymbol{\eta}$ is the combined effect of the additive white Gaussian RX noise and quantization noise that has covariance matrix,  $\mathbf{R}_{\eta} \in \mathbb{C}^{N_\textrm{s} \times N_\textrm{s}}$, given by, 
\begin{equation}\label{eq:R_eta}
\mathbf{R}_{\eta}\!=\!\mathbf{W}^H\mathbf{H} \mathbf{F}_{\textrm{RF}} \mathbf{C}_{\epsilon\textrm{T}} \mathbf{F}_{\textrm{RF}}^H \mathbf{H}^H \mathbf{W}\!+\!\mathbf{W}_\textrm{BB}^H\mathbf{C}_{\epsilon\textrm{R}}\mathbf{W}_\textrm{BB}\!+\!\sigma_\textrm{n}^2 \mathbf{W}^H \mathbf{W}.
\end{equation}
In the following sections, we discuss the joint optimization solution to compute the optimal DAC/ADC bit resolution matrices and the optimal precoder/combiner matrices.

\section{Joint DAC Bit Allocation and A/D Hybrid Precoding Design}
Let us consider a point-to-point MIMO system with a linear quantization model. We define the EE as the ratio of the information rate $R$, i.e. SE, and the total consumed power $P$ \cite{zapponeFTCIT2015} as:
\begin{equation}\label{eq:ee_problem}
 EE \triangleq \frac{R}{P} \,\, \textrm{(bits/Hz/J)}.
\end{equation}
For the given point-to-point MIMO system, the SE is defined as,
\begin{align}\label{eq:rate}
R \!\triangleq\! \log_2 \left \vert \mathbf{I}_{N_\textrm{s}} \!+\! \frac{\mathbf{R}^{-1}_\eta}{N_\textrm{s}} \mathbf{W}^H\mathbf{H}\mathbf{F}  \mathbf{F}^H \mathbf{H}^H \mathbf{W} \right \vert \,\, \textrm{(bits/s/Hz)}, 
\end{align}
where $\mathbf{F} = \mathbf{F}_\textrm{RF} \mathbf{\Delta}_\textrm{TX} \mathbf{F}_\textrm{BB}$ and $\mathbf{W} = \mathbf{W}_\textrm{RF} \mathbf{\Delta}_\textrm{RX} \mathbf{W}_\textrm{BB}$. 

Similar to the power model at the TX in \cite{aryanICC2019}, the total consumed power for the system is expressed as:
\begin{equation}\label{eq:power_total}
P \triangleq P_\textrm{TX}(\mathbf{F}_\textrm{RF}, \mathbf{\Delta}_\textrm{TX}, \mathbf{F}_\textrm{BB}) + P_\textrm{RX}(\mathbf{\Delta}_\textrm{RX}) \,\, \textrm{(W)},
\end{equation}
where the power consumption at the TX is as follows:
\begin{align}\label{eq:tx_power}
P_\textrm{TX}(\mathbf{F}_\textrm{RF}, \mathbf{\Delta}_\textrm{TX}, \mathbf{F}_\textrm{BB}) = &  \textrm{tr}(\mathbf{F}\mathbf{F}^H) + P_\textrm{DT}(\boldsymbol{\Delta}_\textrm{TX}) + N_\textrm{T}P_\textrm{T} 
+ N_\textrm{T}L_\textrm{T}P_\textrm{PT} + P_\textrm{CT} \,\,\textrm{(W)},
\end{align}
where $P_\textrm{PT}$ is the power per phase shifter, $P_\textrm{T}$ is the power per antenna element, $P_\textrm{DT}(\boldsymbol{\Delta}_\textrm{TX})$ is the power associated with the total quantization operation at the TX, and following \eqref{eq:tx_delta_distorsion_rf} and \cite{orhanITA2015}, we have 
\begin{equation}\label{eq:power_dac}
P_\textrm{DT}(\boldsymbol{\Delta}_\textrm{TX}) \!=\! P_\textrm{DAC} \!\sum_{i=1}^{L_{\textrm{T}}} 2^{b_i} \!=\! P_\textrm{DAC} \!\sum_{i=1}^{L_{\textrm{T}}} \left( \frac{\pi \sqrt{3}}{2 (1 \!-\! [\boldsymbol{\Delta}_\textrm{TX}]_{ii}^2)} \right)^{\!\frac{1}{2}} \textrm{(W)},
\end{equation}
where $P_\textrm{DAC}$ is the power consumed per bit in the DAC and $P_\textrm{CT}$ is the power required by all circuit components at the TX. Similarly, the total power consumption at the RX is,
\begin{align}\label{eq:rx_power}
P_\textrm{RX}(\mathbf{\Delta}_\textrm{RX}) \!=\! P_\textrm{DR}(\boldsymbol{\Delta}_\textrm{RX}) \!+\! N_\textrm{R}P_\textrm{R} \!+\! N_\textrm{R}L_\textrm{R}P_\textrm{PR} \!+\! P_\textrm{CR} \,\, \textrm{(W)},
\end{align}
where, at the RX, $P_\textrm{PR}$ is the power per phase shifter, $P_\textrm{R}$ is the power per antenna element, $P_\textrm{DR}$ is the power associated with the total quantization operation, and following \eqref{eq:rx_delta_distorsion_rf} and \cite{orhanITA2015}, we have 
\begin{equation}\label{eq:power_adc}
P_\textrm{DR}(\boldsymbol{\Delta}_\textrm{RX}) \!=\! P_\textrm{ADC} \!\sum_{i=1}^{L_{\textrm{R}}} 2^{b_i} \!=\! P_\textrm{ADC} \!\sum_{i=1}^{L_{\textrm{R}}} \left( \frac{\pi \sqrt{3}}{2 (1 \!-\! [\boldsymbol{\Delta}_\textrm{RX}]_{ii}^2)} \right)^{\!\frac{1}{2}} \textrm{(W)},
\end{equation}
where $P_\textrm{ADC}$ is the power consumed per bit in the ADC and $P_\textrm{CR}$ is the power required by all RX circuit components.

The maximization of EE is given by
\begin{align}
&\underset{\mathbf{F}_\textrm{RF}, \mathbf{\Delta}_\textrm{TX}, \mathbf{F}_\textrm{BB}, \mathbf{W}_\textrm{RF}, \mathbf{\Delta}_\textrm{RX}, \mathbf{W}_\textrm{BB}}{\textrm{max}}
\frac{R(\mathbf{F}_\textrm{RF}, \mathbf{\Delta}_\textrm{TX}, \mathbf{F}_\textrm{BB}, \mathbf{W}_\textrm{RF}, \mathbf{\Delta}_\textrm{RX}, \mathbf{W}_\textrm{BB})}{P_\textrm{TX}(\mathbf{F}_\textrm{RF}, \mathbf{\Delta}_\textrm{TX}, \mathbf{F}_\textrm{BB}) + P_\textrm{RX}(\mathbf{\Delta}_\textrm{RX})} \nonumber \\
\text{ subject to } & \mathbf{F}_\textrm{RF} \in \mathcal{F}^{N_\textrm{T} \times L_\textrm{T}}, 
\mathbf{\Delta}_\textrm{TX} \in \mathcal{D}_\textrm{TX}^{L_\textrm{T} \times L_\textrm{T}}, 
\mathbf{W}_\textrm{RF} \in \mathcal{W}^{N_\textrm{R} \times L_\textrm{R}}, 
\mathbf{\Delta}_\textrm{RX} \in \mathcal{D}_\textrm{RX}^{L_\textrm{R} \times L_\textrm{R}},
\end{align}
when the SE $R$ is given by \eqref{eq:rate} and the power $P$ in \eqref{eq:power_total}. The problem to be addressed involves a fractional cost function that both the numerator and the denominator parts are non-convex functions of the optimizing variables. Furthermore the optimization problem involves non-convex constraint sets. Thus, it is in general a very difficult problem to be addressed. It is interesting that the corresponding problem for a fully digital transceiver that admits a much simpler form is in general intractable due to the coupling of the TX-RX design \cite{palomarJSAC2006}. To that end, we start by decoupling the TX-RX design problem.   

Let us first express the EE maximization problem in the following relaxed form:
\begin{align}
\underset{\mathbf{F}_\textrm{RF}, \mathbf{\Delta}_\textrm{TX}, \mathbf{F}_\textrm{BB}, \mathbf{W}_\textrm{RF}, \mathbf{\Delta}_\textrm{RX}, \mathbf{W}_\textrm{BB}}{\textrm{min}}
&-{R(\mathbf{F}_\textrm{RF}, \mathbf{\Delta}_\textrm{TX}, \mathbf{F}_\textrm{BB}, \mathbf{W}_\textrm{RF}, \mathbf{\Delta}_\textrm{RX}, \mathbf{W}_\textrm{BB})} \nonumber \\
&+\gamma_\textrm{T}{P_\textrm{TX}(\mathbf{F}_\textrm{RF}, \mathbf{\Delta}_\textrm{TX}, \mathbf{F}_\textrm{BB}) + \gamma_\textrm{R}P_\textrm{RX}(\mathbf{\Delta}_\textrm{RX})} \nonumber \\
\text{ subject to } & \mathbf{F}_\textrm{RF} \in \mathcal{F}^{N_\textrm{T} \times L_\textrm{T}}, 
\mathbf{\Delta}_\textrm{TX} \in \mathcal{D}_\textrm{TX}^{L_\textrm{T} \times L_\textrm{T}}, 
\mathbf{W}_\textrm{RF} \in \mathcal{W}^{N_\textrm{R} \times L_\textrm{R}}, 
\mathbf{\Delta}_\textrm{RX} \in \mathcal{D}_\textrm{RX}^{L_\textrm{R} \times L_\textrm{R}},
\label{eq:se0}
\end{align}
where the parameters $\gamma_\textrm{T} \in (0,\gamma_\textrm{T}^{max}] \subset \mathbb{R}^+$ and $\gamma_\textrm{R} \in (0,\gamma_\textrm{R}^{max}] \subset \mathbb{R}^+$ are introducing a trade-off between the achieved rate and the power consumption at the TX's and the RX's side, respectively. Such an approach has been used in the past to tackle fractional optimization problems \cite{dinkelbach}. In the concave/convex case, the equivalence of the relaxed problem with the original fractional one is theoretically established. Unfortunately, a similar result for the case considered in the present paper is not easy to be derived due to the complexity of the addressed problem. Thus, in the present paper, we rely on line search methods in order to optimally tune these parameters.  

Having simplified the original problem, we may now proceed by temporally decoupling the designs at the TX's and the RX's side. Under the assumption that the RX can perform optimal nearest-neighbor decoding based on the received signals, the optimal precoding matrices are designed such that the mutual information achieved by Gaussian signaling over the wireless channel is maximized \cite{ayachTWC2014}. The mutual information is given by
\begin{align}
I \!\triangleq\! \log_2 \left \vert \mathbf{I}_{N_\textrm{s}} \!+\! \frac{\mathbf{Q}_{\eta'}^{-1}}{N_\textrm{s}} \mathbf{H}\mathbf{F}  \mathbf{F}^H \mathbf{H}^H \right \vert \,\, \textrm{(bits/s/Hz)}, 
\label{eq:mi}
\end{align}
where again $\mathbf{F} = \mathbf{F}_\textrm{RF} \mathbf{\Delta}_\textrm{TX} \mathbf{F}_\textrm{BB}$ and
and $\mathbf{Q}_{\eta'}$ is the covariance matrix of the sum of noise and transmit quantization noise variables, i.e. $\eta' = \mathbf{F}_{\textrm{RF}}  \boldsymbol{\epsilon}_\textrm{TX} + \mathbf{n}$, given by
\begin{equation}\label{eq:Q_eta}
\mathbf{Q}_{\eta'}\!=\!\mathbf{F}_{\textrm{RF}} \mathbf{C}_{\epsilon\textrm{T}} \mathbf{F}_{\textrm{RF}}^H\!+\!\sigma_\textrm{n}^2 \mathbf{I}_{N_\textrm{R}}.
\end{equation}

Based on \eqref{eq:se0}-\eqref{eq:mi}, the precoding matrices may be derived as the solution to the following optimization problem:
\begin{align*}
(\mathcal{P}_\textrm{1T}): \hspace{5pt} \min_{\mathbf{F}_\textrm{RF}, \mathbf{\Delta}_\textrm{TX}, \mathbf{F}_\textrm{BB}} \,\, -{I(\mathbf{F}_\textrm{RF}, \mathbf{\Delta}_\textrm{TX}, \mathbf{F}_\textrm{BB})}+\gamma_T{P_\textrm{TX}(\mathbf{F}_\textrm{RF}, \mathbf{\Delta}_\textrm{TX}, \mathbf{F}_\textrm{BB})},  \\
\textrm{ subject to } \mathbf{F}_\textrm{RF} \in \mathcal{F}^{N_\textrm{T} \times L_\textrm{T}}, 
\mathbf{\Delta}_\textrm{TX} \in \mathcal{D}_\textrm{TX}^{L_\textrm{T} \times L_\textrm{T}},
\end{align*}

Now provided that the optimal precoding matrix $\mathbf{F}^\star = \mathbf{F}_\textrm{RF}^\star \mathbf{\Delta}_\textrm{TX}^\star \mathbf{F}_\textrm{BB}^\star$ is derived from solving $(\mathcal{P}_\textrm{1T})$, we can plug in these resulted precoding matrices in the cost function of \eqref{eq:se0} resulting in an optimization problem dependent only on the decoder matrices at the RX's side, defined as,
\begin{align}
(\mathcal{P}_\textrm{1R}): \hspace{5pt}  \underset{\mathbf{W}_\textrm{RF}, \mathbf{\Delta}_\textrm{RX}, \mathbf{W}_\textrm{BB}}{\textrm{min}}
&-{\tilde{R}(\mathbf{W}_\textrm{RF}, \mathbf{\Delta}_\textrm{RX}, \mathbf{W}_\textrm{BB})} +\gamma_RP_{RX}(\mathbf{\Delta}_\textrm{RX}) \nonumber \\
\text{ subject to } & \mathbf{W}_\textrm{RF} \in \mathcal{W}^{N_\textrm{R} \times L_\textrm{R}}, 
\mathbf{\Delta}_\textrm{RX} \in \mathcal{D}_\textrm{RX}^{L_\textrm{R} \times L_\textrm{R}},
\label{eq:se}
\end{align}
where ${\tilde{R}(\mathbf{W}_\textrm{RF}, \mathbf{\Delta}_\textrm{RX}, \mathbf{W}_\textrm{BB})} = R(\mathbf{F}_\textrm{RF}^\star, \mathbf{\Delta}_\textrm{TX}^\star, \mathbf{F}_\textrm{BB}^\star, \mathbf{W}_\textrm{RF}, \mathbf{\Delta}_\textrm{RX}, \mathbf{W}_\textrm{BB})$.

Thus, the precoding and decoding matrices can be derived as the solutions to the two decoupled problems $(\mathcal{P}_\textrm{1T})-(\mathcal{P}_\textrm{1R})$ above. In the following subsections, the solutions to these problems are developed. We start first with the development of the solution to TX's side one $(\mathcal{P}_\textrm{1T})$ and then the solution for the RX's side $(\mathcal{P}_\textrm{1R})$ counterpart follows.  

\subsection{Problem Formulation at the TX}
Focusing on the TX side, we seek the bit resolution matrix $\boldsymbol{\Delta}_\textrm{TX}$ and the hybrid precoding matrices $\mathbf{F}_\textrm{RF}$, $\mathbf{F}_\textrm{BB}$ that solve $(\mathcal{P}_\textrm{1T})$.  The set $\mathcal{D}_\textrm{TX}$ represents the finite states of the quantizer and is defined as,
\[\mathcal{D}_\textrm{TX} \!=\! \left\{\!\mathbf{\Delta}_\textrm{TX} \in \mathbb{R}^{L_\textrm{T} \times L_\textrm{T}} \big| m \le [\mathbf{\Delta}_\textrm{TX}]_{ii} \le M \hspace{1mm} \forall \hspace{1mm} i = 1,..., L_\textrm{T}\!\right\}.\] 
Note that $P_\textrm{TX}(\mathbf{F}_\textrm{RF}, \mathbf{\Delta}_\textrm{TX}, \mathbf{F}_\textrm{BB}) > 0$, as defined in \eqref{eq:tx_power}, since the power required by all circuit components is always larger than zero, i.e., $P_\textrm{CP}>0$.

Since dealing with the part of the cost function of $(\mathcal{P}_\textrm{1T})$ that involves the mutual information expression is a difficult task due to the perplexed form of the latter, we adopt the approach in \cite{ayachTWC2014} where the maximization of the mutual information $I$ can be approximated by finding the minimum Euclidean distance of the hybrid precoder to the one of the fully digital transceiver for the full-bit resolution sampling case, denoted by $\mathbf{F}_\textrm{DBF}$, i.e., $\Vert\mathbf{F}_\textrm{DBF} - \mathbf{F}_\textrm{RF}\mathbf{\Delta}_\textrm{TX}\mathbf{F}_\textrm{BB}\Vert_F^2$ \cite{ayachTWC2014}. Therefore, motivated by the previous, $(\mathcal{P}_\textrm{1T})$ can be approximated to finding the solution of the following problem:
\begin{align*}
\hspace{-5pt}(\mathcal{P}_\textrm{2}): \hspace{5pt} &\min_{\mathbf{F}_\textrm{RF},\mathbf{\Delta}_\textrm{TX},\mathbf{F}_\textrm{BB}} \frac{1}{2}\Vert\mathbf{F}_\textrm{DBF}-\mathbf{F}_\textrm{RF}\mathbf{\Delta}_\textrm{TX}\mathbf{F}_\textrm{BB}\Vert_F^2 + \gamma_\textrm{T} P_\textrm{TX}(\mathbf{F}), \nonumber \\
& 
\textrm{subject to}
\hspace{1mm} \mathbf{F}_\textrm{RF} \in \mathcal{F}^{N_\textrm{T} \times L_\textrm{T}},
\mathbf{\Delta}_\textrm{TX} \in \mathcal{D}_\textrm{TX}^{L_\textrm{T} \times L_\textrm{T}}.
\end{align*}
For a point-to-point MIMO system the optimal $\mathbf{F}_\textrm{DBF}$ is given by $\mathbf{F}_\textrm{DBF} = \mathbf{V}(\mathbf{P})^{\frac{1}{2}}$ where the orthonormal matrix $\mathbf{V} \in \mathbb{C}^{N_\textrm{R} \times N_\textrm{T}}$ is derived via the channel matrix singular value decomposition (SVD), i.e. $\mathbf{H} = \mathbf{U}\mathbf{\Sigma}\mathbf{V}^H$ and $\mathbf{P}$ is a diagonal power allocation matrix with real positive diagonal entries derived by the so-called ``water-filling algorithm'' \cite{tse2004}. 
 
Problem $(\mathcal{P}_\textrm{2})$ is still very difficult to address as it is non-convex due to the non-convex cost function that involves the product of three matrix variables and non-convex constraints. In the next section, an efficient algorithmic solution based on the ADMM is proposed.

\subsection{Proposed ADMM Solution at the TX}
In the following we develop an iterative procedure for solving $(\mathcal{P}_\textrm{2})$ based on the ADMM approach \cite{ADMM}. This method is a variant of the standard augmented Lagrangian method that uses partial updates (similar to the Gauss-Seidel method for the solution of linear equations) to solve constrained optimization problems. While it is mainly known for its good performance for a number of convex optimization problems, recently it has been successfully applied to non-convex matrix factorization as well \cite{ADMM,TSINOSNC1,TSINOSNC2}. Motivated by this, in the following ADMM based solutions are developed that are tailored for the non-convex matrix factorization problem $(\mathcal{P}_\textrm{2})$.  

We first transform $(\mathcal{P}_\textrm{2})$ into a form that can be addressed via ADMM. By using the auxiliary variable $\mathbf{Z}$, $(\mathcal{P}_\textrm{2})$ can be written as:
\begin{align*}
\hspace{-2pt}(\mathcal{P}_\textrm{3}): \hspace{2pt}  &\min_{\substack{\mathbf{Z},\mathbf{F}_\textrm{RF}, \mathbf{\Delta}_\textrm{TX},\mathbf{F}_\textrm{BB}}} \frac{1}{2}\Vert\mathbf{F}_\textrm{DBF}-\mathbf{Z}\Vert_F^2 + \mathds{1}_{\mathcal{F}^{N_\textrm{T} \times L_\textrm{T}}}\{\mathbf{F}_\textrm{RF}\} 
+ \mathds{1}_{\mathcal{D}_\textrm{TX}^{L_\textrm{T} \times L_\textrm{T}}}\{\mathbf{\Delta}_\textrm{TX}\} + \gamma_\textrm{T} P_\textrm{TX}(\mathbf{F}), \\
& \hspace{20mm} \textrm{subject to} \hspace{1mm} \mathbf{Z} = \mathbf{F}_\textrm{RF}\mathbf{\Delta}_\textrm{TX}\mathbf{F}_\textrm{BB}.
\end{align*} 
Problem $(\mathcal{P}_\textrm{3})$ formulates the A/D hybrid precoder matrix design as a matrix factorization problem. That is, the overall precoder $\mathbf{Z}$ is sought so that it minimizes the Euclidean distance to the optimal, fully digital precoder $\mathbf{F}_\textrm{DBF}$ while supporting decomposition into three factors: the analog precoder matrix $\mathbf{F}_\textrm{RF}$, the DAC bit resolution matrix $\boldsymbol{\Delta}_\textrm{TX}$ and the digital precoder matrix $\mathbf{F}_\textrm{BB}$. The augmented Lagrangian function of $(\mathcal{P}_\textrm{3})$ is given by
\begin{align}\label{EQ:tx_AUGLAN} 
\mathcal{L}(\mathbf{Z},\mathbf{F}_\textrm{RF},\mathbf{\Delta}_\textrm{TX},\mathbf{F}_\textrm{BB},\mathbf{\Lambda}) =& \frac{1}{2}\Vert\mathbf{F}_\textrm{DBF}\!-\!\mathbf{Z}\Vert^2_F  \!+\! \mathds{1}_{\mathcal{F}^{N_\textrm{T} \times L_\textrm{T}}}\{\mathbf{F}_\textrm{RF}\}+ 
\!\mathds{1}_{\mathcal{D}_\textrm{TX}^{L_\textrm{T} \times L_\textrm{T}}}\{{\mathbf{\Delta}_\textrm{TX}}\} \nonumber\\  
&\!+\! \frac{\alpha}{2}\Vert\mathbf{Z}\!+\!\mathbf{\Lambda}/\alpha\! - \!\mathbf{F}_\textrm{RF}\mathbf{\Delta}_\textrm{TX}\mathbf{F}_\textrm{BB}\Vert_F^2 
\!+\! \gamma_\textrm{T} P_\textrm{TX}(\mathbf{F}),
\end{align}
where $\alpha$ is a scalar penalty parameter and $\mathbf{\Lambda} \in \mathbb{C}^{N_\textrm{T}\times L_\textrm{T}}$ is the Lagrange Multiplier matrix. According to the ADMM approach \cite{ADMM}, the solution to $(\mathcal{P}_3)$ is derived by the following iterative steps where $n$ denotes the iteration index:
\allowdisplaybreaks
\begin{align}
&\hspace{0pt}(\mathcal{P}_\textrm{3A}): \hspace{0pt} \mathbf{Z}_{(n)} = \arg \min_{\mathbf{Z}} \mathcal{L}(\mathbf{Z},\mathbf{F}_{\textrm{RF}(n-1)},\mathbf{\Delta}_{\textrm{TX}(n-1)},
\mathbf{F}_{\textrm{BB}(n-1)}, \mathbf{\Lambda}_{(n-1)}), \nonumber \\
&\hspace{0pt}(\mathcal{P}_\textrm{3B}): \hspace{0pt} \mathbf{F}_{\textrm{RF}(n)} = \arg \min_{\mathbf{F}_\textrm{RF}}\mathcal{L}(\mathbf{Z}_{(n)},\mathbf{F}_\textrm{{RF}},\mathbf{\Delta}_{\textrm{TX}(n-1)}, 
\mathbf{F}_{\textrm{BB}(n-1)},\mathbf{\Lambda}_{(n-1)}), \nonumber \\
&\hspace{0pt}(\mathcal{P}_\textrm{3C}): \hspace{0pt} \mathbf{\Delta}_{\textrm{TX}(n)} = \arg \min_{\mathbf{\Delta}_\textrm{TX}}\mathcal{L}(\mathbf{Z}_{(n)},\mathbf{F}_{\textrm{RF}(n)},\mathbf{\Delta}_\textrm{TX},
\mathbf{F}_{\textrm{BB}(n-1)},\mathbf{\Lambda}_{(n-1)})\!+\!\gamma_\textrm{T} P_\textrm{TX}(\mathbf{F}), \nonumber\\
&\hspace{0pt}(\mathcal{P}_\textrm{3D}): \hspace{0pt} \mathbf{F}_{\textrm{BB}(n)} = \arg \min_{\mathbf{F}_\textrm{BB}}\mathcal{L}(\mathbf{Z}_n,\mathbf{F}_{\textrm{RF}(n)},\mathbf{\Delta}_{\textrm{TX}(n)},
\mathbf{F}_\textrm{BB},\mathbf{\Lambda}_{(n-1)}), \nonumber \\
& \ \mathbf{\Lambda}_{(n)} = \mathbf{\Lambda}_{(n-1)} + \alpha\left(\mathbf{Z}_{(n)}-\mathbf{F}_{\textrm{RF}(n)}\mathbf{\Delta}_{\textrm{TX}(n)}\mathbf{F}_{\textrm{BB}(n)}\right).
\label{EQ:LnGR} 
\end{align}

In order to apply the ADMM iterative procedure, we have to solve the optimization problems $(\mathcal{P}_\textrm{3A})$-$(\mathcal{P}_\textrm{3D})$. We may start from problem $(\mathcal{P}_\textrm{3A})$ which can be written as follows:  
\begin{align}
 \hspace{0pt} (\mathcal{P}_\textrm{3A}'):\hspace{5pt} \mathbf{Z}_{(n)} = \arg \min_{\mathbf{Z}}  \frac{1}{2}\Vert(1+\alpha)\mathbf{Z}-\mathbf{F}_\textrm{DBF}+\mathbf{\Lambda}_{(n-1)}- 
 \alpha\mathbf{F}_{\textrm{RF}(n-1)}\mathbf{\Delta}_{\textrm{TX}(n-1)}\mathbf{F}_{\textrm{BB}(n-1)}\Vert_F^2. \nonumber
\end{align}
Problem $(\mathcal{P}_\textrm{3A}')$ can be directly solved by equating the gradient of the augmented Lagrangian \eqref{EQ:tx_AUGLAN} w.r.t. $\mathbf{Z}$ being set to zero. Therefore, we have
\begin{align}
\label{EQ:Z_SOL}
\mathbf{Z}_{(n)} \!=\! \frac{1}{\alpha \!+\! 1}\Big(\!\mathbf{F}_\textrm{DBF}\! -\! \mathbf{\Lambda}_{(n-1)} \!+\! \alpha\mathbf{F}_{\textrm{RF}(n-1)}\mathbf{\Delta}_{\textrm{TX}(n-1)}\mathbf{F}_{\textrm{BB}(n-1)}\!\Big).
\end{align}

We may now proceed to solve $(\mathcal{P}_\textrm{3B})$ which can be written in the following simplified form by keeping only the terms of the augmented Lagrangian that are dependent on $\mathbf{F}_\textrm{RF}$:
\begin{align}
 \hspace{0pt} (\mathcal{P}_\textrm{3B}'):\hspace{5pt}  \mathbf{F}_{\textrm{RF}(n)} = \arg \min_{\mathbf{F}_\textrm{RF}}  \mathds{1}_{\mathcal{F}^{N_\textrm{T} \times L_\textrm{T}}}\{\mathbf{F}_\textrm{RF}\}+ \frac{\alpha}{2}
  \Vert\mathbf{Z}_{(n)}+\mathbf{\Lambda}_{(n-1)}/\alpha- \mathbf{F}_\textrm{RF}\mathbf{\Delta}_{\textrm{TX}(n-1)}\mathbf{F}_{\textrm{BB}(n-1)}\Vert_F^2. \nonumber
\end{align}
The solution to problem $(\mathcal{P}_{3B}')$ does not admit a closed form and thus, it is approximated by solving the unconstrained problem and then projecting onto the set $\mathcal{F}^{N_\textrm{T} \times L_\textrm{T}}$, i.e.,
\begin{align}\label{EQ:FRF_SOL}
\mathbf{F}_{\textrm{RF}(n)} =\Pi_\mathcal{F}\Big\{\left(\mathbf{\Lambda}_{(n-1)}+\alpha\mathbf{Z}_{(n)}\right)\mathbf{F}_{\textrm{BB}(n-1)}^H\mathbf{\Delta}_{\textrm{TX}(n-1)}^H 
\left(\alpha\mathbf{\Delta}_{\textrm{TX}(n-1)}\mathbf{F}_{\textrm{BB}(n-1)}\mathbf{F}_{\textrm{BB}(n-1)}^H\mathbf{\Delta}_{\textrm{TX}(n-1)}^H\right)^{-1}\Big\},
\end{align}
where ${\Pi}_{\mathcal{F}}$ projects the solution onto the set $\mathcal{F}$. This is computed by solving the following optimization problem \cite{AGL}:

\begin{align}
(\mathcal{P}_\textrm{3B}^{''}): \hspace{25pt} &\min_{\mathbf{A}_{\mathcal{F}}}\|\mathbf{A}_{\mathcal{F}}-\mathbf{A}\|_F^2, \textrm{subject to} \ \mathbf{A}_{\mathcal{F}} \in \mathcal{F}, \nonumber
\end{align}
where $\mathbf{A}$ is an arbitrary matrix and $\mathbf{A}_{\mathcal{F}}$ is its projection onto the set $\mathcal{F}$. The solution to $(\mathcal{P}_\textrm{3B}^{''})$ is given by the phase of the complex elements of $\mathbf{A}$. Thus, for $\mathbf{A}_\mathcal{F}  = \Pi_\mathcal{F}\{\mathbf{A}\}$ we have
\begin{align}
\mathbf{A}_\mathcal{F}(x,y) = \begin{cases}
0, \ &\mathbf{A}(x,y) = 0 \\
\frac{\mathbf{A}(x,y)}{\left|\mathbf{A}(x,y)\right|}, \ &\mathbf{A}(x,y) \neq 0 
\end{cases},  
\label{EQ:PROJ_F}
\end{align}
where $\mathbf{A}_\mathcal{F}(x,y)$ and $\mathbf{A}(x,y)$ are the elements at the $x$th row-$y$th column of matrices $\mathbf{A}_\mathcal{F}$ and $\mathbf{A}$, respectively. While, this is an approximate solution, it turns out that it behaves remarkably well, as verified in the simulation results of Section V. This is due to the interesting property that ADMM is observed to converge even in cases where the alternating minimization steps are not carried out exactly \cite{ADMM}. There are theoretical results that support this statement \cite{BER,GOL}, though an exact analysis for the case considered here is beyond the scope of this paper. 
\begin{algorithm}[t]
    \caption{Proposed ADMM Solution for the A/D Hybrid Precoder Design}
  \begin{algorithmic}[1]
    \STATE \textbf{Initialize:} $\mathbf{Z}$, ${\mathbf{F}_\textrm{RF}}$, $\mathbf{\Delta}_\textrm{TX}$, ${\mathbf{F}_\textrm{BB}}$ with random values, $\mathbf{\Lambda}$ with zeros, $\alpha=1$ and $n=1$
    \WHILE{The termination criteria of \eqref{EQ:TERM} are not met or $n \leq N_\textrm{max}$}
      	\STATE Update $\mathbf{Z}_{(n)}$ using solution \eqref{EQ:Z_SOL}, \\ \hspace{10mm} $\mathbf{F}_{\textrm{RF}(n)}$ using solution \eqref{EQ:FRF_SOL}, \\ \hspace{10mm}  $\mathbf{\Delta}_{\textrm{TX}(n)}$ by solving ($\mathcal{P}_\textrm{3C}^{''}$) using CVX \cite{cvx},\\ \hspace{10mm} $\mathbf{F}_{\textrm{BB}(n)}$ using solution \eqref{EQ:FBB_SOL}, and \\
      	\hspace{10mm} update $\mathbf{\Lambda}_{(n)}$ using solution \eqref{EQ:LnGR}.
      	\STATE $n \gets n+1$
      \ENDWHILE
 \RETURN $\mathbf{F}_\textrm{RF}^\star$, $\mathbf{\Delta}_\textrm{TX}^\star$, $\mathbf{F}_\textrm{BB}^\star$
  \end{algorithmic}
\end{algorithm} 

In a similar manner, $(\mathcal{P}_\textrm{3C})$ may be re-written as,
\begin{align*}
(\mathcal{P}_\textrm{3C}'): \mathbf{\Delta}_{\textrm{TX}(n)} =& \arg \min_{\mathbf{\Delta}_\textrm{TX}}  \mathds{1}_{\mathcal{D}_\textrm{TX}^{L_\textrm{T} \times L_\textrm{T}}}\{{\mathbf{\Delta}_\textrm{TX}}\}+ \frac{\alpha}{2}\Vert\mathbf{Z}_{(n)}+ 
\mathbf{\Lambda}_{(n-1)}/\alpha - \mathbf{F}_{\textrm{RF}(n)}\mathbf{\Delta}_\textrm{TX}\mathbf{F}_{\textrm{BB}(n-1)}\Vert_F^2 \nonumber \\
&+ \gamma_\textrm{T} P_\textrm{TX}(\mathbf{F}).
\end{align*}
To solve the above problem, we can write:
\begin{align}
(\mathcal{P}_\textrm{3C}^{''}): \hspace{0pt} \mathbf{\Delta}_{\textrm{TX}(n)} \!=\! &\arg \min_{\mathbf{\Delta}_\textrm{TX}}\Vert\mathbf{y}_\textrm{c}\!-\!\mathbf{\Psi}_\textrm{T}\textrm{vec}(\mathbf{\Delta}_\textrm{TX})\Vert_2^2 \!+\!\gamma_\textrm{T} P_\textrm{TX}(\mathbf{F}), \nonumber \\
&\textrm{ subject to } \mathbf{\Delta}_\textrm{TX} \in \mathcal{D}_\textrm{TX}, \nonumber
\end{align}
The minimization problem in $(\mathcal{P}_\textrm{3C}^{''})$ consists of $\mathbf{y}_c = \textrm{vec}(\textbf{Z}_n + \Lambda_{n-1}/\alpha)$, $\mathbf{\Psi}_\textrm{T} = \mathbf{F}_{\textrm{BB}(n-1)} \otimes \mathbf{F}_{\textrm{RF}(n)}$ ($\otimes$ being the Khatri-Rao product) and is solved using CVX \cite{cvx}.

The solution of problem $(\mathcal{P}_\textrm{3D})$ may be written in the following form:
\begin{align}
 \hspace{0pt} (\mathcal{P}_\textrm{3D}'):\hspace{5pt} \mathbf{F}_{\textrm{BB}(n)} = \arg \min_{\mathbf{F}_\textrm{BB}} \frac{\alpha}{2}\Vert\mathbf{Z}_{(n)}+\mathbf{\Lambda}_{(n-1)}/\alpha 
 - \mathbf{F}_{\textrm{RF}(n)}\mathbf{\Delta}_{\textrm{TX}(n)}\mathbf{F}_\textrm{BB}\Vert_F^2. \nonumber
\end{align}
It is straightforward to see that the solution for $(\mathcal{P}_\textrm{3D}')$ can be obtained by equating the gradient to zero and solving the resulting equation w.r.t. the matrix variable $\mathbf{F}_\textrm{BB}$, i.e.,
\begin{align}
\label{EQ:FBB_SOL}
\mathbf{F}_{\textrm{BB}(n)} =
\left(\alpha\mathbf{\Delta}_{\textrm{TX}(n)}^H\mathbf{F}_{\textrm{RF}(n)}^H\mathbf{F}_{\textrm{RF}(n)}\mathbf{\Delta}_{\textrm{TX}(n)}\right)^{-1} 
\mathbf{\Delta}_{\textrm{TX}(n)}^H \mathbf{F}_{\textrm{RF}(n)}^H\left(\mathbf{\Lambda}_{(n-1)}+\alpha\mathbf{Z}_{(n)}\right).
\end{align}

Algorithm 1 provides the complete procedure to obtain the optimal analog precoder matrix $\mathbf{F}_\textrm{RF}$, the optimal bit resolution matrix $\mathbf{\Delta}_\textrm{TX}$ and the optimal baseband (or digital) precoder matrix $\mathbf{F}_\textrm{BB}$. It starts the alternating minimization procedure by initializing the entries of the matrices $\mathbf{Z}$, ${\mathbf{F}_\textrm{RF}}$, $\mathbf{\Delta}_\textrm{TX}$, ${\mathbf{F}_\textrm{BB}}$ with random values and the entries of the Lagrange multiplier matrix $\mathbf{\Lambda}$ with zeros. For iteration index $n$, $\mathbf{Z}_{(n)}$, ${\mathbf{F}_\textrm{RF}}_{(n)}$, $\mathbf{\Delta}_{\textrm{TX}(n)}$ and ${\mathbf{F}_\textrm{BB}}_{(n)}$ are updated using Step 3 which shows the steps to be used to obtain the matrices. A termination criterion related to either the maximum permitted number of iterations ($N_\textrm{max}$) is considered or the ADMM solution meeting the following criteria is considered:    

{
\setlength{\abovedisplayskip}{6pt}
  \setlength{\belowdisplayskip}{\abovedisplayskip}
  \setlength{\abovedisplayshortskip}{0pt}
  \setlength{\belowdisplayshortskip}{3pt}
  \vspace{-10pt}
\begin{align}
\norm{\mathbf{Z}_{(n)} - \mathbf{Z}_{(n-1)}}_F \leq \epsilon^z \ \& \ 
\Vert\mathbf{Z}_{(n)} - \mathbf{F}_{\textrm{RF}(n)}\mathbf{\Delta}_{\textrm{TX}(n)}\mathbf{F}_{\textrm{BB}(n)}\Vert_F \leq \epsilon^p,
\label{EQ:TERM}
\end{align}  }
\hspace{-10pt} where $\epsilon^z$ and $\epsilon^p$ are the corresponding tolerances. Upon convergence, the number of bits for each DAC is obtained by using \eqref{eq:tx_delta_distorsion_rf} and quantizing to the nearest integer value. The optimal hybrid precoding matrices $\mathbf{F}_\textrm{RF}^\star$, $\mathbf{\Delta}_\textrm{TX}^\star$, $\mathbf{F}_\textrm{BB}^\star$ are obtained at the end of this algorithm. 

\subsubsection*{Computational complexity analysis of Algorithm 1}
When running Algorithm 1, mainly Step 3, while updating $\mathbf{\Delta}_{\textrm{TX}(n)}$ by solving ($\mathcal{P}_\textrm{3C}^{''}$) using CVX, involves multiplication by $\mathbf{\Psi}_\textrm{T}$ whose dimensions are $L_\textrm{T}N_\textrm{T}\times N_\textrm{s} L_\textrm{T}$. In general, the solution of $(\mathcal{P}_\textrm{3C}^{''})$ can be upper-bounded by $\mathcal{O}((L_\textrm{T}^2N_\textrm{T}N_\textrm{s})^3)$ which can be improved significantly by exploiting the structure of $\mathbf{\Psi}_\textrm{T}$.

In the following section, we discuss the joint optimization problem at the RX and the solution to obtain the analog combiner matrix $\mathbf{W}_\textrm{RF}$, the ADC bit resolution matrix $\mathbf{\Delta}_\textrm{RX}$ and the digital combiner matrix $\mathbf{W}_\textrm{BB}$. 

\section{Joint ADC Bit Allocation and A/D Hybrid Combining Optimization}
\subsection{Problem Formulation at the RX}
Let us now move to the derivation of the solution to $(\mathcal{P}_\textrm{1R})$. The set $\mathcal{D}_\textrm{RX}$ represents the finite states of the ADC quantizer and is defined as,
\[\mathcal{D}_\textrm{RX} \!=\! \left\{\!\mathbf{\Delta}_\textrm{RX} \in \mathbb{R}^{L_\textrm{R} \times L_\textrm{R}} \big| m \le [\mathbf{\Delta}_\textrm{RX}]_{ii} \le M \hspace{1mm} \forall \hspace{1mm} i = 1,..., L_\textrm{R}\!\right\}.\]


Due to the perplexed form of the function ${\tilde{R}(\mathbf{W}_\textrm{RF}, \mathbf{\Delta}_\textrm{RX}, \mathbf{W}_\textrm{BB})}$, we follow the same arguments the under of which we approximated $(\mathcal{P}_\textrm{2})$ by $(\mathcal{P}_\textrm{1T})$,  in order to approximate $(\mathcal{P}_\textrm{1R})$ by
\begin{align*}
\hspace{0pt}(\mathcal{P}_\textrm{5}): \hspace{0pt} &\min_{\mathbf{W}_\textrm{RF},\mathbf{\Delta}_\textrm{RX},\mathbf{W}_\textrm{BB}} \frac{1}{2}\Vert\mathbf{W}_\textrm{DBF}\!-\!\mathbf{W}_\textrm{RF}\mathbf{\Delta}_\textrm{RX}\mathbf{W}_\textrm{BB}\Vert_F^2 
\!+\! \gamma_\textrm{R} P_\textrm{RX}(\mathbf{\Delta}_\textrm{RX}),\nonumber \\
& \textrm{subject to}
\hspace{1mm} \mathbf{W}_\textrm{RF} \in \mathcal{W}^{N_\textrm{R} \times L_\textrm{R}}, \mathbf{\Delta}_\textrm{RX} \in \mathcal{D}_\textrm{RX}^{L_\textrm{R} \times L_\textrm{R}},
\end{align*}
where $\mathbf{W}_\textrm{DBF}$ is the optimal solution for the fully digital RX which is given by $\mathbf{W}_\textrm{DBF} = (\tilde{\mathbf{P}})^{\frac{1}{2}}\tilde{\mathbf{U}}$, where $\tilde{\mathbf{U}} \in \mathbb{C}^{N_\textrm{R}\times N_\textrm{s}}$ is the orthonormal singular vector matrix which can be derived by the SVD of the equivalent channel matrix $\tilde{\mathbf{H}} = \mathbf{H}\mathbf{F}^\star = \tilde{\mathbf{U}}\tilde{\mathbf{\Sigma}}\tilde{\mathbf{V}}^H$, and $\tilde{\mathbf{P}}$ is diagonal power allocation matrix. 
Problem $(\mathcal{P}_\textrm{5})$ is also non-convex due to the non-convex cost function and non-convex set of constraints, as well, and for its solution an ADMM-based solution similar to the case of $(\mathcal{P}_\textrm{2})$ is derived in the following subsection.


\subsection{Proposed ADMM Solution at the RX}
In the following we develop an iterative procedure for solving $(\mathcal{P}_\textrm{5})$ based on ADMM \cite{ADMM}. We first transform $(\mathcal{P}_\textrm{5})$ into an amenable form. By using the auxiliary variable $\mathbf{Z}$, $(\mathcal{P}_\textrm{5})$ can be written as:
\begin{align*}
\hspace{-2pt}(\mathcal{P}_\textrm{6}): \hspace{2pt}  \min_{\substack{\mathbf{Z},\mathbf{W}_\textrm{RF}, \mathbf{\Delta}_\textrm{RX},\mathbf{W}_\textrm{BB}}} &\frac{1}{2}\Vert\mathbf{W}_\textrm{DBF}-\mathbf{Z}\Vert_F^2 + \mathds{1}_{\mathcal{W}^{N_\textrm{R} \times L_\textrm{R}}}\{\mathbf{W}_\textrm{RF}\} 
+ \mathds{1}_{\mathcal{D}_\textrm{RX}^{L_\textrm{R} \times L_\textrm{R}}}\{\mathbf{\Delta}_\textrm{RX}\} + \gamma_\textrm{R} P_\textrm{RX}(\mathbf{\Delta}_\textrm{RX}), \\
\textrm{ subject to } & \mathbf{Z} = \mathbf{W}_\textrm{RF}\mathbf{\Delta}_\textrm{RX}\mathbf{W}_\textrm{BB}.
\end{align*} 
Problem $(\mathcal{P}_\textrm{6})$ formulates the A/D hybrid combiner matrix design as a matrix factorization problem. That is, the overall combiner $\mathbf{Z}$ is sought so that it minimizes the Euclidean distance to the optimal, fully digital combiner $\mathbf{W}_\textrm{DBF}$ while supporting the decomposition into the analog combiner matrix $\mathbf{W}_\textrm{RF}$, the quantization error matrix $\boldsymbol{\Delta}_\textrm{RX}$ and the digital combiner matrix $\mathbf{W}_\textrm{BB}$. The augmented Lagrangian function of $(\mathcal{P}_\textrm{6})$ is given by
\begin{align}
\mathcal{L}(\mathbf{Z},\mathbf{W}_\textrm{RF},\mathbf{\Delta}_\textrm{RX},\mathbf{W}_\textrm{BB}, \mathbf{\Lambda}) =&  \frac{1}{2}\lVert\mathbf{W}_\textrm{DBF}-\mathbf{Z}\rVert^2_F+\mathds{1}_{\mathcal{W}^{N_\textrm{R} \times L_\textrm{R}}}\{\mathbf{W}_\textrm{RF}\} 
+\mathds{1}_{\mathcal{D}_\textrm{RX}^{L_\textrm{R} \times L_\textrm{R}}}\{{\mathbf{\Delta}_\textrm{RX}}\} \nonumber\\ 
&+\frac{\alpha}{2}\lVert\mathbf{Z}+\mathbf{\Lambda}/\alpha - \mathbf{W}_\textrm{RF}\mathbf{\Delta}_\textrm{RX}\mathbf{W}_\textrm{BB}\rVert_F^2
+\gamma_\textrm{R} P_\textrm{RX}(\mathbf{\Delta}_\textrm{RX}),
\label{EQ:AUGLAN} 
\end{align}
where $\alpha$ is a scalar penalty parameter and $\mathbf{\Lambda} \in \mathbb{C}^{N_\textrm{R}\times L_\textrm{R}}$ is the Lagrange Multiplier matrix. According to the ADMM approach \cite{ADMM}, the solution to  $(\mathcal{P}_\textrm{6})$ is derived by the following iterative steps:
\allowdisplaybreaks
\begin{align}
&\hspace{0pt}(\mathcal{P}_\textrm{6A}): \hspace{0pt} \mathbf{Z}_{(n)} = \arg \min_{\mathbf{Z}}  \frac{1}{2}\lVert(1+\alpha)\mathbf{Z}-\mathbf{W}_\textrm{DBF}+\mathbf{\Lambda}_{(n-1)} 
-\alpha\mathbf{W}_{\textrm{RF}(n-1)}\mathbf{\Delta}_{\textrm{RX}(n-1)}\mathbf{W}_{\textrm{BB}(n-1)}\rVert_F^2, \nonumber\\
%
&\hspace{0pt}(\mathcal{P}_\textrm{6B}): \hspace{0pt} \mathbf{W}_{\textrm{RF}(n)} = \arg \min_{\mathbf{W}_\textrm{RF}}  \mathds{1}_{\mathcal{W}^{N_\textrm{R} \times L_\textrm{R}}}\{\mathbf{W}_\textrm{RF}\}\!+\! \frac{\alpha}{2}
\norm{\mathbf{Z}_{(n)}\!+\!\mathbf{\Lambda}_{(n-1)}/\alpha\!-\! \mathbf{W}_\textrm{RF}\mathbf{\Delta}_{\textrm{RX}(n-1)}\mathbf{W}_{\textrm{BB}(n-1)}}_F^2, \nonumber\\
%
&\hspace{0pt}(\mathcal{P}_\textrm{6C}): \hspace{0pt} \mathbf{\Delta}_{\textrm{RX}(n)} = \arg \min_{\mathbf{\Delta}_\textrm{RX}}\Vert\mathbf{y}_\textrm{c}-\mathbf{\Psi}_\textrm{R}\textrm{vec}(\mathbf{\Delta}_\textrm{RX})\Vert_2^2 
+\gamma_\textrm{R} P_\textrm{RX}(\mathbf{\Delta}_\textrm{RX}) \textrm{ subject to } \mathbf{\Delta}_\textrm{RX} \in \mathcal{D}_\textrm{RX},\nonumber\\
%
&\hspace{0pt}(\mathcal{P}_\textrm{6D}): \hspace{0pt} \mathbf{W}_{\textrm{BB}(n)} = \arg \min_{\mathbf{W}_\textrm{BB}} \frac{\alpha}{2}\Vert\mathbf{Z}_{(n)}+\mathbf{\Lambda}_{(n-1)}/\alpha 
-\mathbf{W}_{\textrm{RF}(n)}\mathbf{\Delta}_{\textrm{RX}(n)}\mathbf{W}_\textrm{BB}\Vert_F^2, \nonumber\\
\label{EQ:rx_LnGR} 
&\mathbf{\Lambda}_{(n)} = \mathbf{\Lambda}_{(n-1)} + \alpha\left(\mathbf{Z}_{(n)}-{\mathbf{W}_\textrm{RF}}_{(n)}\mathbf{\Delta}_{\textrm{RX}(n)}{\mathbf{W}_\textrm{BB}}_{(n)}\right),
\end{align}
where $n$ denotes the iteration index, $\mathbf{y}_\textrm{c}\!=\! \textrm{vec}(\textbf{Z}_{(n)}\!+\! \mathbf{\Lambda}_{(n-1)}/\alpha)$ and $\mathbf{\Psi}_\textrm{R}\!=\! {\mathbf{W}_\textrm{BB}}_{(n-1)}\! \otimes \! {\mathbf{W}_\textrm{RF}}_{(n)}$ ($\otimes$ is the Khatri-Rao product). 

\begin{algorithm}[t]
    \caption{Proposed ADMM Solution for the A/D Hybrid Combiner Design}
  \begin{algorithmic}[1]
    \STATE \textbf{Initialize:} $\mathbf{Z}$, ${\mathbf{W}_\textrm{RF}}$, $\mathbf{\Delta}_\textrm{RX}$, ${\mathbf{W}_\textrm{BB}}$ with random values, $\mathbf{\Lambda}$ with zeros, $\alpha=1$ and $n=1$
    \WHILE{$n \leq N_\textrm{max}$}
      	\STATE Update $\mathbf{Z}_{(n)}$ using solution \eqref{eq:rx_Z_sol}, \\ \hspace{10mm} $\mathbf{W}_{\textrm{RF}(n)}$ using solution \eqref{eq:rx_Wrf_sol}, \\ \hspace{10mm}  $\mathbf{\Delta}_{\textrm{RX}(n)}$ by solving ($\mathcal{P}_\textrm{6C}$) using CVX \cite{cvx},\\ \hspace{10mm} $\mathbf{W}_{\textrm{BB}(n)}$ using solution \eqref{eq:rx_Wbb_sol}, and \\
      	\hspace{10mm} update $\mathbf{\Lambda}_{(n)}$ using solution \eqref{EQ:rx_LnGR}.\\
      	\STATE $n \gets n+1$
      \ENDWHILE
    \RETURN $\mathbf{W}_\textrm{RF}^\star$, $\mathbf{\Delta}_\textrm{RX}^\star$, $\mathbf{W}_\textrm{BB}^\star$
  \end{algorithmic}
\end{algorithm}

We solve the optimization problems $(\mathcal{P}_\textrm{6A})$-$(\mathcal{P}_\textrm{6D})$ in a similar way to the derivations in Section III for the TX. The solution for $\mathbf{Z}_{(n)}$ is:
\begin{equation}\label{eq:rx_Z_sol}
\mathbf{Z}_{(n)} =\frac{1}{\alpha + 1} \big(\mathbf{W}_\textrm{DBF} - \mathbf{\Lambda}_{(n-1)} +  \alpha {\mathbf{W}_\textrm{RF}}_{(n-1)}\mathbf{\Delta}_{\textrm{RX}(n-1)} {\mathbf{W}_\textrm{BB}}_{(n-1)}\big).
\end{equation}
The equation for ${\mathbf{W}_\textrm{RF}}_{(n)}$ is as follows:
\begin{align}\label{eq:rx_Wrf_sol}
{\mathbf{W}_\textrm{RF}}_{(n)} =  \Pi_\mathcal{W}\Big\{\big(\mathbf{\Lambda}_{(n-1)}+\alpha\mathbf{Z}_{(n)}\big){\mathbf{W}_\textrm{BB}}_{(n-1)}^H\mathbf{\Delta}_{\textrm{RX}(n-1)}^H \nonumber \\
\big\{\alpha\mathbf{\Delta}_{\textrm{RX}(n-1)}{\mathbf{W}_\textrm{BB}}_{(n-1)}{\mathbf{W}_\textrm{BB}}_{(n-1)}^H\mathbf{\Delta}_{\textrm{RX}(n-1)}^H\big\}^{-1}\Big\}.     
\end{align}
The solution to $\mathbf{\Delta}_{\textrm{RX}(n)}$ is obtained by solving $(\mathcal{P}_\textrm{6C})$ using CVX \cite{cvx}. The matrix ${\mathbf{W}_\textrm{BB}}_{(n)}$ is obtained as follows:
\begin{align}\label{eq:rx_Wbb_sol}
{\mathbf{W}_\textrm{BB}}_{(n)} = \big\{\alpha\mathbf{\Delta}_{\textrm{RX}(n)}^H{\mathbf{W}_\textrm{RF}}_{(n)}^H{\mathbf{W}_\textrm{RF}}_{(n)}\mathbf{\Delta}_{\textrm{RX}(n)}\big\}^{-1}\mathbf{\Delta}_{\textrm{RX}(n)}^H 
{\mathbf{W}_\textrm{RF}}_{(n)}^H\big(\mathbf{\Lambda}_{(n-1)}+\alpha\mathbf{Z}_{(n)}\big).
\end{align}

Algorithm 2 provides the complete procedure to obtain $\mathbf{W}_\textrm{RF}$, $\mathbf{\Delta}_\textrm{RX}$ and $\mathbf{W}_\textrm{BB}$. It starts by initializing the entries of the matrices $\mathbf{Z}$, ${\mathbf{W}_\textrm{RF}}$, $\mathbf{\Delta}_\textrm{RX}$, ${\mathbf{W}_\textrm{BB}}$ with random values and the entries of the Lagrange multiplier matrix $\mathbf{\Lambda}$ with zeros. For iteration index $n$, $\mathbf{Z}_{(n)}$, ${\mathbf{W}_\textrm{RF}}_{(n)}$,  $\mathbf{\Delta}_{\textrm{RX}(n)}$, ${\mathbf{W}_\textrm{BB}}_{(n)}$ are updated at each iteration step by using the solution in \eqref{eq:rx_Z_sol}, \eqref{eq:rx_Wrf_sol}, solving $(\mathcal{P}_\textrm{6C})$ using CVX, \eqref{eq:rx_Wbb_sol} and \eqref{EQ:rx_LnGR}, respectively. The operator ${\Pi}_{\mathcal{W}}$ projects the solution onto the set $\mathcal{W}$. This procedure is identical to problem ($\mathcal{P}_\textrm{3B}^{''}$) in Section III, except that the set $\mathcal{W}$ replaces $\mathcal{F}$.
A termination criterion is defined using a maximum number of iterations ($N_\textrm{max}$) or a fidelity criterion similar to \eqref{EQ:TERM}.
Upon convergence, the number of bits for each ADC is obtained by using \eqref{eq:rx_delta_distorsion_rf} and quantizing to the nearest integer value. The optimal hybrid combining matrices $\mathbf{W}_\textrm{RF}^\star$, $\mathbf{\Delta}_\textrm{RX}^\star$, $\mathbf{W}_\textrm{BB}^\star$ are obtained at the end of this algorithm.

\subsubsection*{Computational complexity analysis of Algorithm 2}
Similar to Algorithm 1 for the TX, the complexity of the solution of $(\mathcal{P}_\textrm{6C})$ can be upper-bounded by $\mathcal{O}((L_\textrm{R}^2N_\textrm{R}N_\textrm{s})^3)$ which can be improved significantly by exploiting the structure of $\mathbf{\Psi}_\textrm{R}$.

Once the optimal DAC and ADC bit resolution matrices, i.e., $\mathbf{\Delta}_\textrm{TX}$ and $\mathbf{\Delta}_\textrm{RX}$, and optimal hybrid precoding and combining matrices, i.e., $\mathbf{F}_\textrm{RF}$, $\mathbf{F}_\textrm{BB}$ and $\mathbf{W}_\textrm{RF}$, $\mathbf{W}_\textrm{BB}$, are obtained then they can be plugged into \eqref{eq:rate} and \eqref{eq:power_total} to obtain the maximum EE in \eqref{eq:ee_problem}.
In the next section, we discuss the simulation results based on the proposed solution at the TX and the RX, and comparison with existing benchmark techniques.

\section{Simulation Results}
In this section, we evaluate the performance of the proposed ADMM solution using computer simulation results. All the results have been averaged over 1000 Monte-Carlo realizations. For comparison with the proposed ADMM solution, we consider following benchmark techniques:
\subsubsection{Digital beamforming with 8-bit resolution}
We consider the conventional fully digital beamforming architecture, where the number of RF chains at the TX/RX is equal to the number of TX/RX antennas, i.e., $L_\textrm{T} = N_\textrm{T}$ and $L_\textrm{R} = N_\textrm{R}$. In terms of the resolution sampling, we consider full-bit resolution, i.e., $M = 8$-bit, which represents the best case from the achievable SE perspective.

\subsubsection{A/D Hybrid beamforming with 1-bit and 8-bit resolutions}
We also consider a A/D hybrid beamforming architecture with $L_\textrm{T} < N_\textrm{T}$ and $L_\textrm{R} < N_\textrm{R}$, for two cases of DAC/ADC bit resolution: a) 1-bit resolution which usually shows reasonable EE performance, and b) 8-bit resolution which usually shows high SE results. 

\subsubsection{Brute force with A/D hybrid beamforming}
We also implement an exhaustive search approach as an upper bound for EE maximization called brute force (BF), based on \cite{ranziSAC2016}. Firstly the EE problem is split into TX and RX optimization problems similar to those for the proposed ADMM approach. Then it makes a search over all the possible DAC and ADC bit resolutions in the range of $[m, M]$ associated with the each RF chain from $1$ to $L_\textrm{T}$ and $1$ to $L_\textrm{R}$ at the TX and the RX, respectively. It then finds the best EE out of all the possible cases and chooses the corresponding optimal resolution for each DAC and ADC. This method provides the best possible EE performance and serves as upper bound for EE maximization by the ADMM approach.

\begin{table}
    \centering
    \bgroup
\def\arraystretch{1}
\begin{subtable}{1.0\textwidth}
\centering
\begin{tabular}{ |c|c|}
 \hline
 \textbf{Power Terms} & \textbf{Values} \\
 \hline
Power per bit in the DAC/ADC & $P_\textrm{DAC}  = P_\textrm{ADC}= 100$ mW \\
 \hline
 Circuit power at the TX/RX & $P_\textrm{CT} = P_\textrm{CR}= 10$ W \\
 \hline
 Power per phase shifter at the TX/RX & $P_\textrm{PT} = P_\textrm{PR}= 10$ mW \\
 \hline
 Power per antenna at the TX/RX & $P_\textrm{T} = P_\textrm{R}$ = $100$ mW \\
 \hline
\end{tabular}
\vspace{2mm}
\caption{Typical values of the power terms \cite{rappa2} used in (16) and (18).}
\vspace{2mm}
\end{subtable} \\
\begin{subtable}{1.0\textwidth}
\centering
\begin{tabular}{ |c|c|} 
   \hline
 \textbf{System Parameters} & \textbf{Values} \\
 \hline
   Number of clusters   & $N_\textrm{cl} = 2$ \\
  \hline
  Number of rays   & $N_\textrm{ray} = 3$ \\
  \hline
  Number of TX antennas   & $N_\textrm{T} = 32$   \\
  \hline
  Number of RX antennas   & $N_\textrm{R} = 5$ \\
  \hline
   Number of TX/RX RF chains & $L_\textrm{T} = L_\textrm{R} = 5$ \\
  \hline
  Number of data streams   & $N_\textrm{s} = L_\textrm{T} = 5$ \\
  \hline
    Bit resolution range   & $[m, M] = [1, 8]$ \\
  \hline
   Maximum number of ADMM iterations & $N_\textrm{max} = 20$ \\
  \hline
   Maximum TX/RX trade-off parameter   & $\gamma_\textrm{T}^{max} = 0.1$; $\gamma_\textrm{R}^{max}$ = 1 \\
  \hline
 \end{tabular}
 \vspace{2mm}
 \centering
 \caption{System parameter values.}
\end{subtable}
\egroup
\caption{Summary of the simulation parameter values.}
\label{tab:Table1}
\vspace{-4mm}
\end{table}

\subsubsection*{Complexity comparison with the BF approach} The proposed ADMM solution has lower complexity than the upper bound BF approach because the BF technique involves a search over all the possible DAC/ADC bit resolutions while the proposed ADMM solution directly optimizes the number of bits at each DAC/ADC. We constrain the number of RF chains $L_\textrm{T}=L_\textrm{R}=5$ for the BF approach due to the high complexity order which is $\mathcal{O}(M^{L_\textrm{T}})$ and $\mathcal{O}(M^{L_\textrm{R}})$ at the TX and the RX, respectively.

\begin{figure}[t]
\centering
\begin{subfigure}{0.495\textwidth}
\centering
    \includegraphics[width=1.115\textwidth]{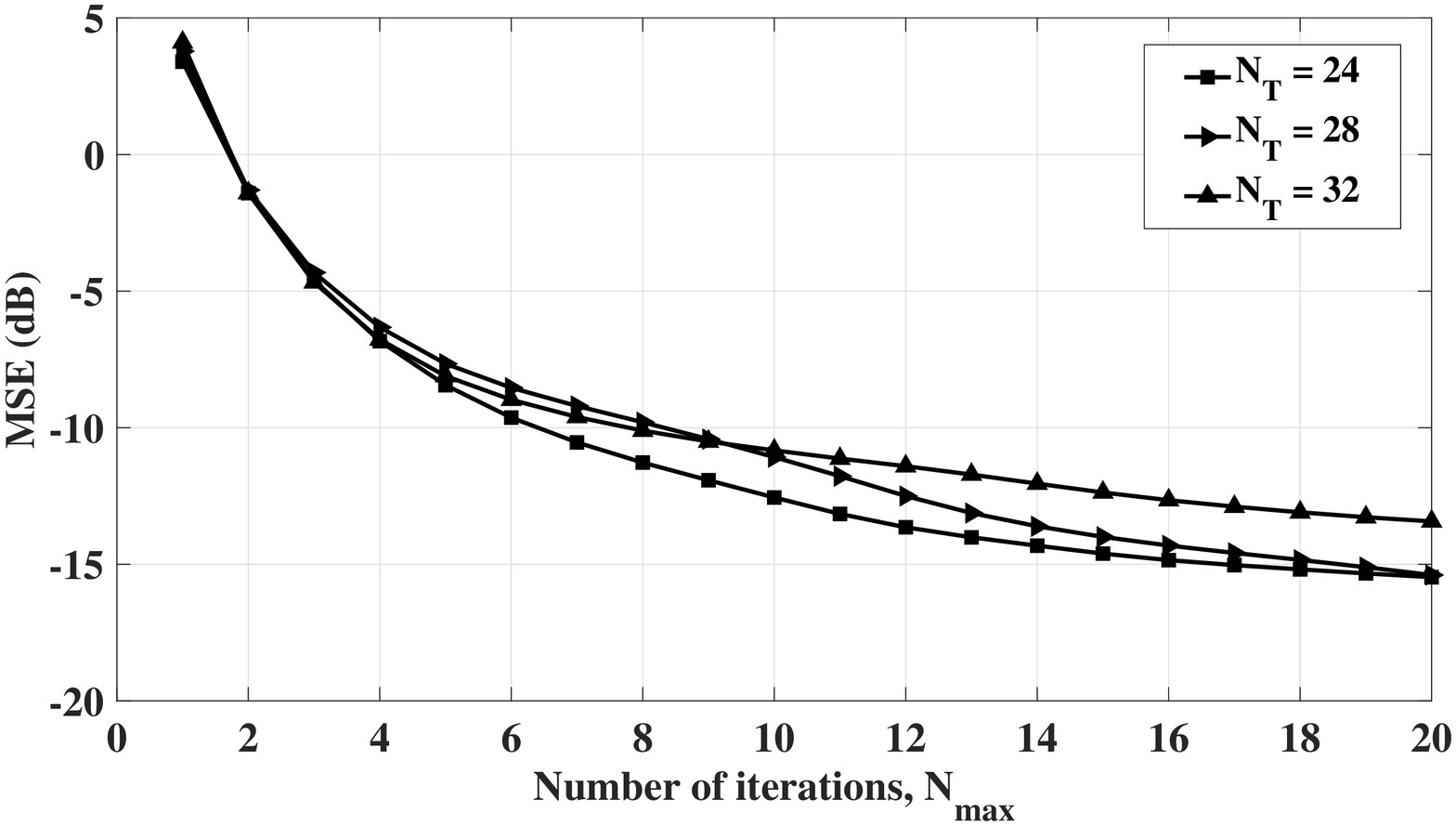}
		\caption{At the TX for different $N_\textrm{T}$ at $\gamma_\textrm{T} = 0.001$.}
\end{subfigure}
 \begin{subfigure}{0.495\textwidth}
 \centering
     \includegraphics[width=1.115\textwidth]{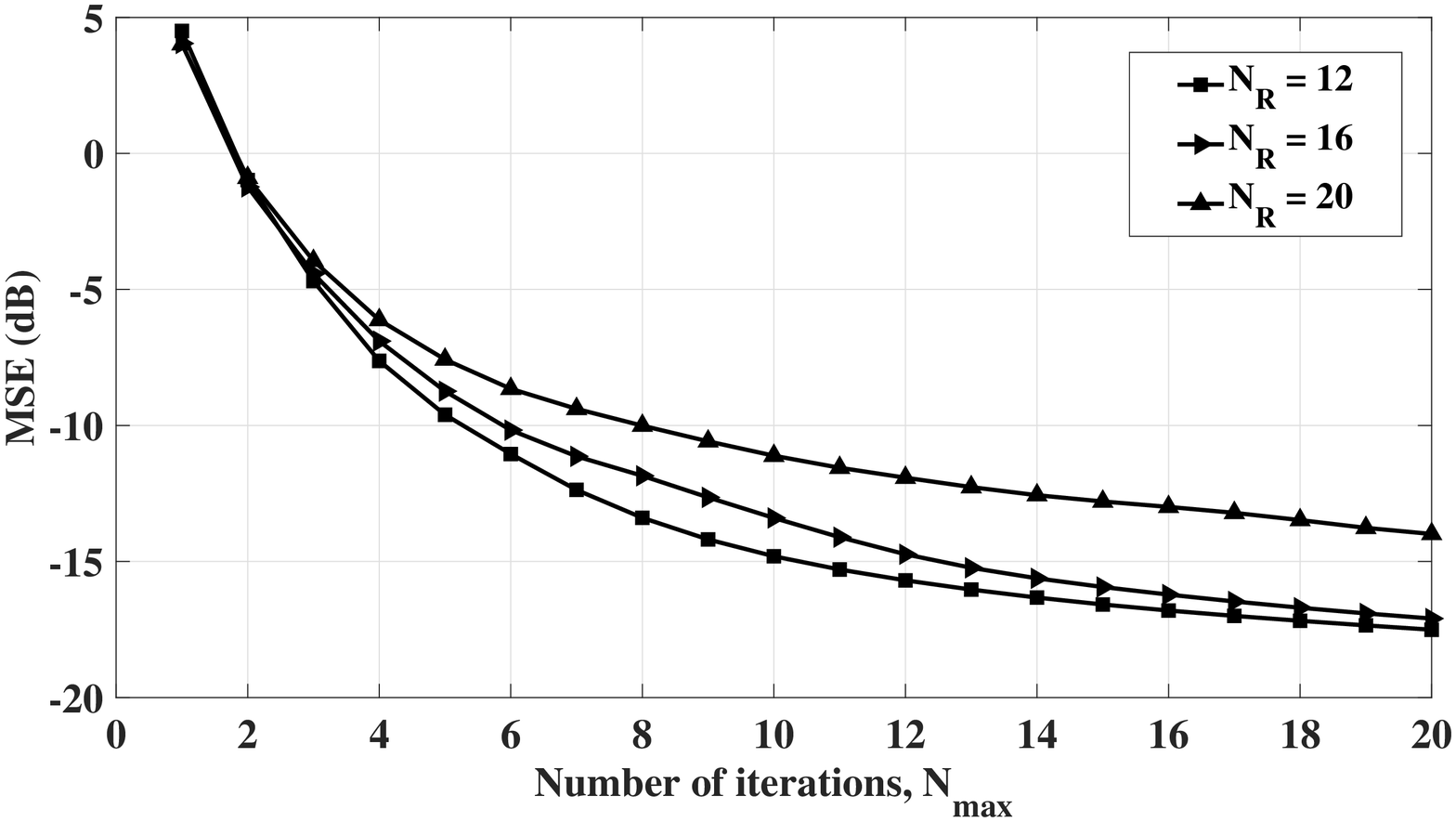}
 		\caption{At the RX for different $N_\textrm{R}$ at $\gamma_\textrm{R} = 0.5$.}
 \end{subfigure}
\caption{Convergence of the proposed ADMM solution at the TX and the RX.}
\vspace{-4mm}
\end{figure}


\subsubsection*{System setup}
Table 1 summarizes the simulation values used for the system and power terms, and in addition, we consider $\alpha=1$ and $\sigma^2_{\alpha, i}$ = $1$. The azimuth angles of departure and arrival are computed with uniformly distributed mean angles, and each cluster follows a Laplacian distribution about the mean angle. The antenna elements in the ULA are spaced by distance $d = \lambda/2$. The signal-to-noise ratio (SNR) is given by the inverse of the noise variance, i.e., $1/\sigma_\textrm{n}^2$. The transmit vector $\mathbf{x}$ is composed of the normalized i.i.d. Gaussian symbols. Under this assumption the covariance matrix of $\mathbf{x}$ is an identity matrix.

\subsubsection*{Convergence of the proposed ADMM solution}
Figs. 2 (a) and 2 (b) show the convergence of the ADMM solution at the TX and the RX as proposed in Algorithm 1 and Algorithm 2, respectively, to obtain the optimal bit resolution at each DAC/ADC and the corresponding optimal precoder/combiner matrices. It can be observed from Fig. 2 (a) that the proposed solution converges rapidly within 16 iterations and the normalized mean square error (NMSE) at the TX, $\norm{\mathbf{F}_\textrm{DBF} - \mathbf{F}_{\textrm{RF}(N_\textrm{max})}\mathbf{\Delta}_{\textrm{TX}(N_\textrm{max})}\mathbf{F}_{\textrm{BB}(N_\textrm{max})}}_F^2/\norm{\mathbf{F}_\textrm{DBF}}_F^2$, goes as low as -15 dB. Similarly, in Fig. 2 (b), the proposed solution again converges rapidly and the NMSE at the RX, $\norm{\mathbf{W}_\textrm{DBF} - \mathbf{W}_{\textrm{RF}(N_\textrm{max})}\mathbf{\Delta}_{\textrm{RX}(N_\textrm{max})}\mathbf{W}_{\textrm{BB}(N_\textrm{max})}}_F^2/\norm{\mathbf{W}_\textrm{DBF}}_F^2$, goes as low as $-17$ dB. A lower number of TX/RX antennas shows lower NMSE for a given number of iterations as expected, since fewer parameters are required to be estimated.

\begin{figure}[t]
 \centering \includegraphics[width=0.85\textwidth]{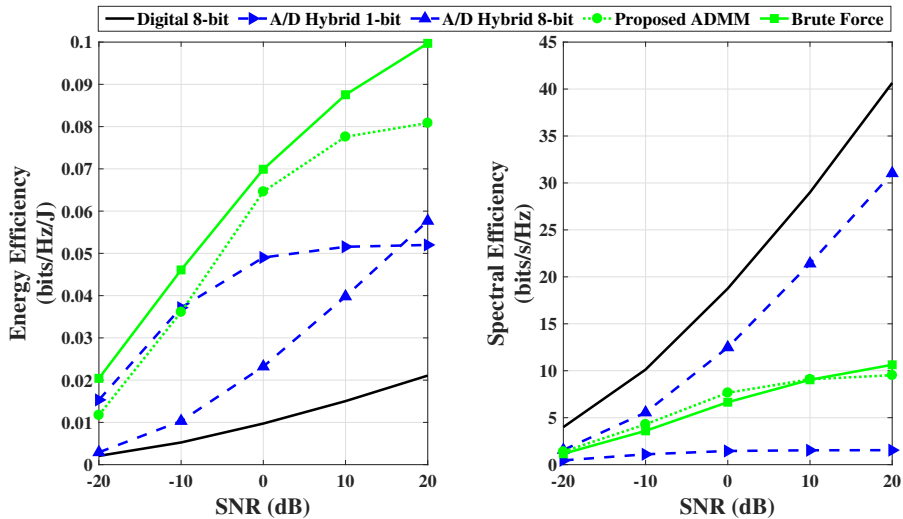}
 		\caption{EE and SE performance w.r.t. SNR at  $\gamma_\textrm{T}=0.001$ and $\gamma_\textrm{R}=0.5$.}
 \end{figure}
 
 \begin{figure}[t]  
 \centering \includegraphics[width=0.85\textwidth]{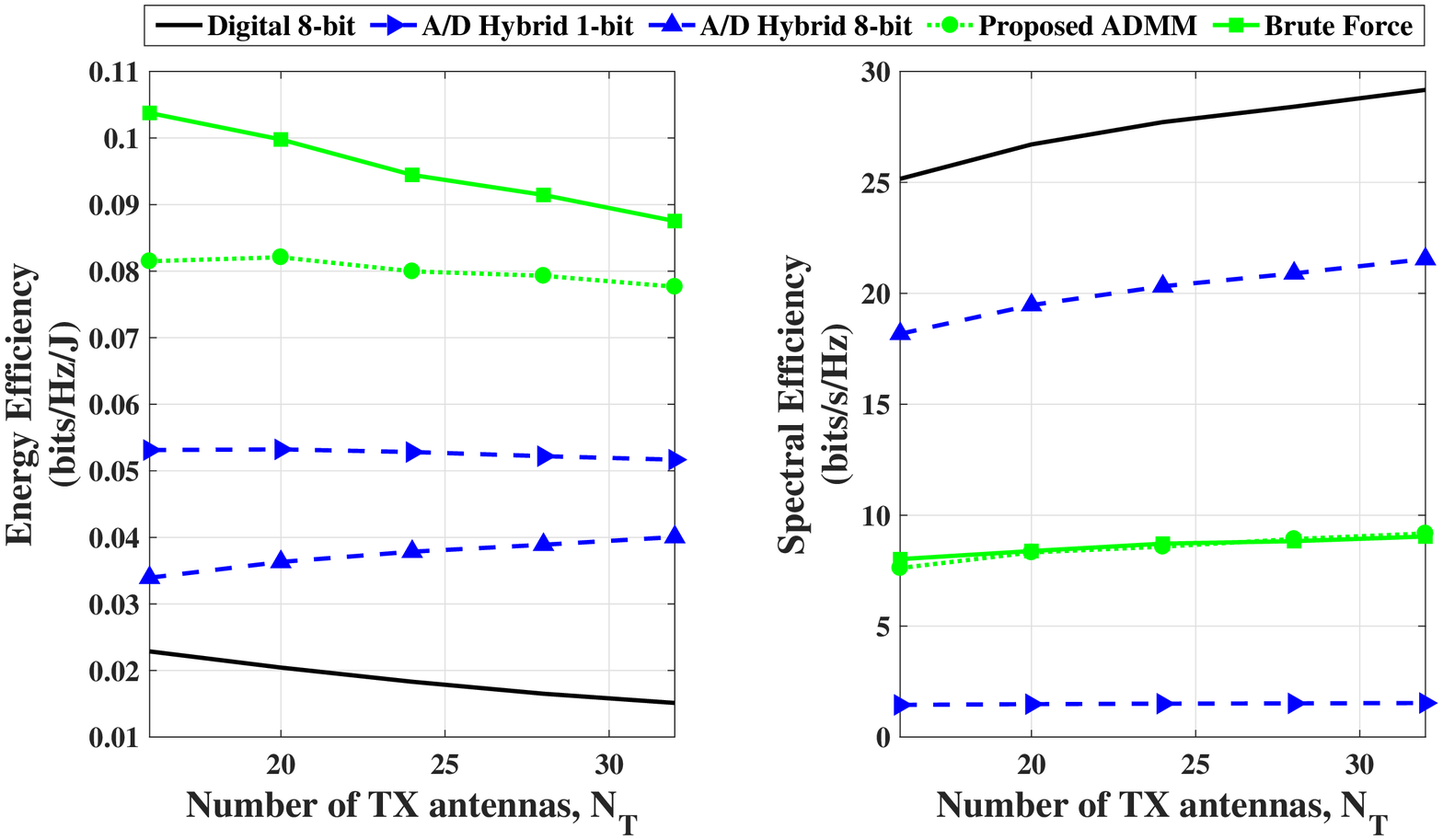}
 		\caption{EE and SE performance w.r.t. $N_\textrm{T}$ at SNR = $10$ dB, $\gamma_\textrm{T}=0.001$ and $\gamma_\textrm{R}=0.5$.}
 		\vspace{-4mm}
 \end{figure}
 

\begin{figure}[t] 
\centering \includegraphics[width=0.85\textwidth]{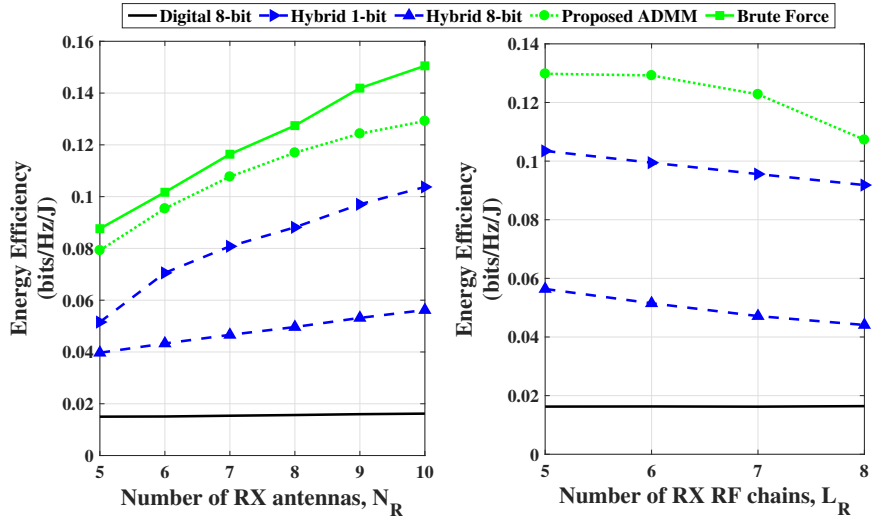}
 		\caption{EE performance w.r.t. $N_\textrm{R}$ and $L_\textrm{R}$ at SNR = $10$ dB, $\gamma_\textrm{T}=0.001$ and $\gamma_\textrm{R}=0.5$.}
 \end{figure}

\begin{figure}[t] 
\centering \includegraphics[width=0.85\textwidth]{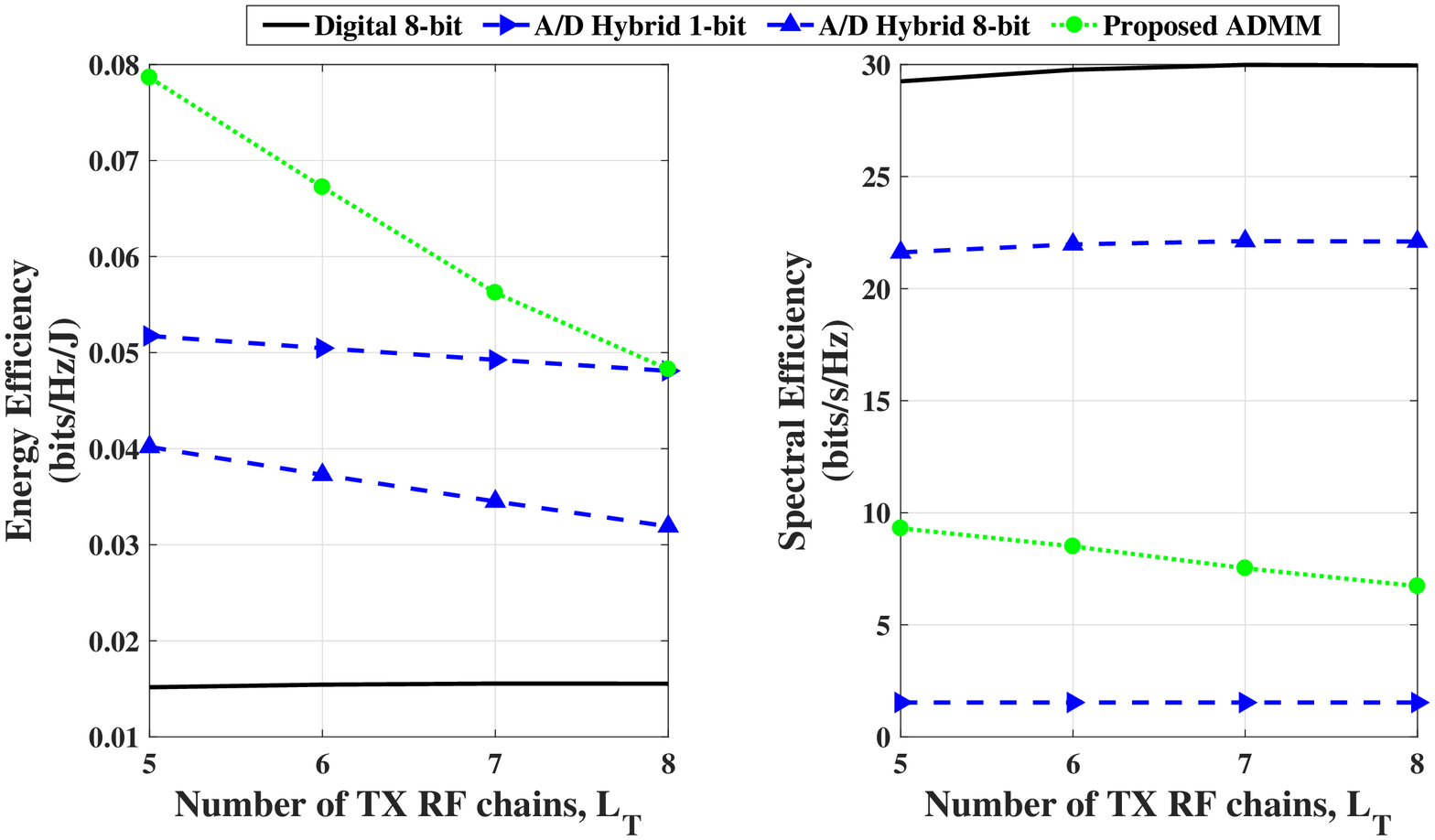}
 		\caption{EE and SE performance w.r.t. $L_\textrm{T}$ at SNR = $10$ dB, $\gamma_\textrm{T}=0.001$ and $\gamma_\textrm{R}=0.5$.}
 		\vspace{-4mm}
 \end{figure}
 

\begin{figure}[t] 
\centering 
\includegraphics[width=0.85\textwidth]{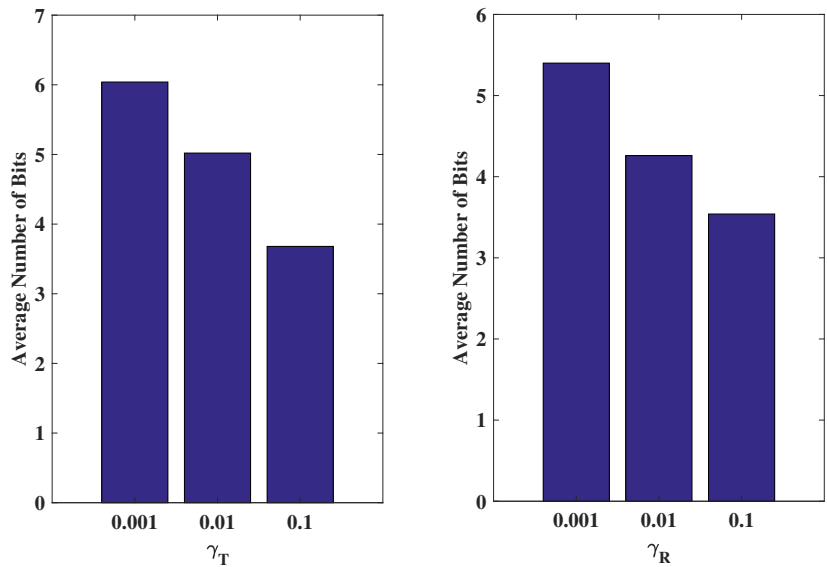}
 		\caption{Average number of bits for proposed ADMM w.r.t. $\gamma_\textrm{T}$ and $\gamma_\textrm{R}$ at the TX and the RX, respectively, at SNR = $10$ dB.}
 \end{figure}

\begin{figure}[t] 
\centering 
\includegraphics[width=0.85\textwidth]{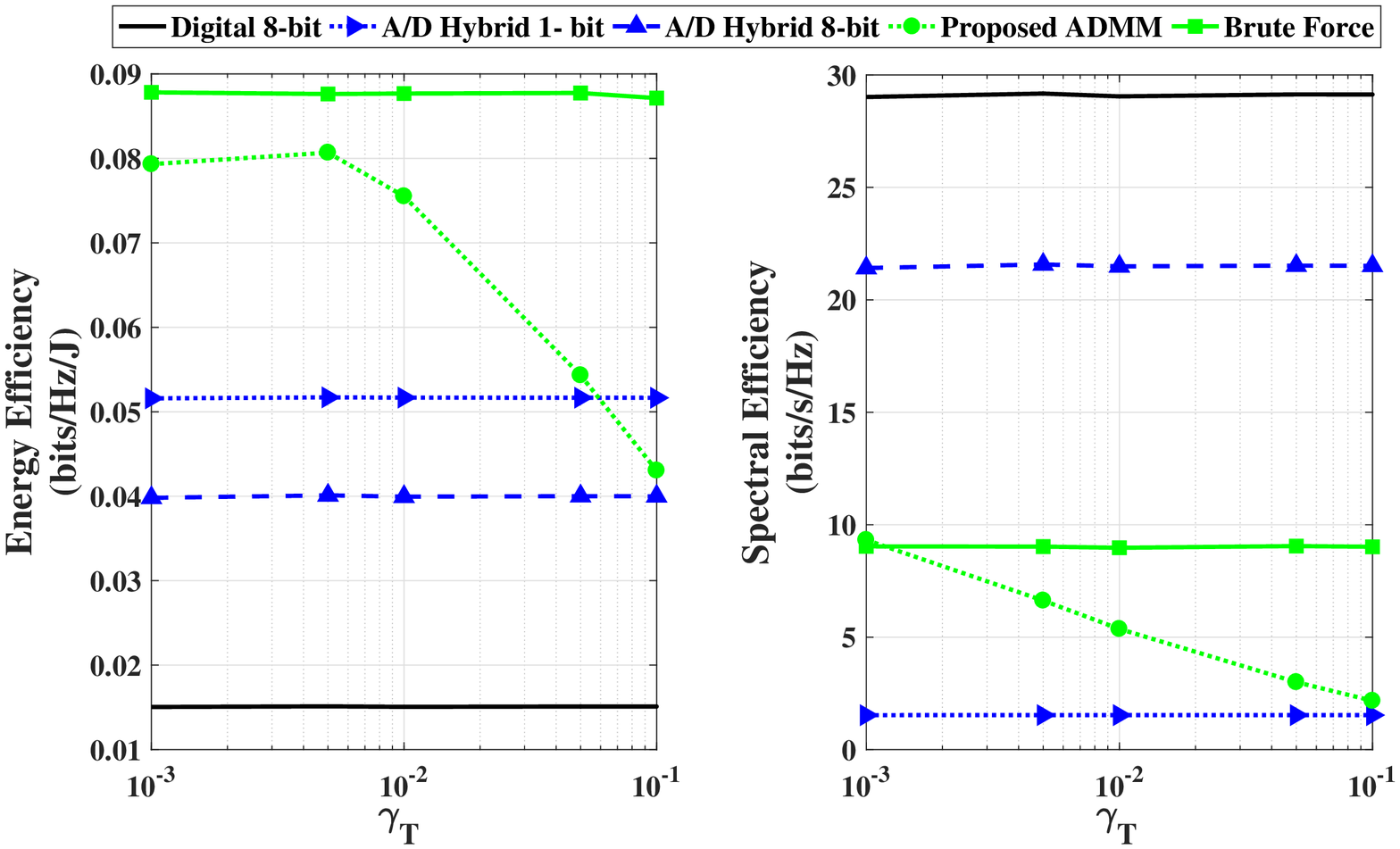}
 		\caption{EE and SE performance w.r.t. $\gamma_\textrm{T}$ at SNR = $10$ dB.}
 		\vspace{-4mm}
 \end{figure}
 
 \begin{figure}[t] 
\centering 
\includegraphics[width=0.85\textwidth]{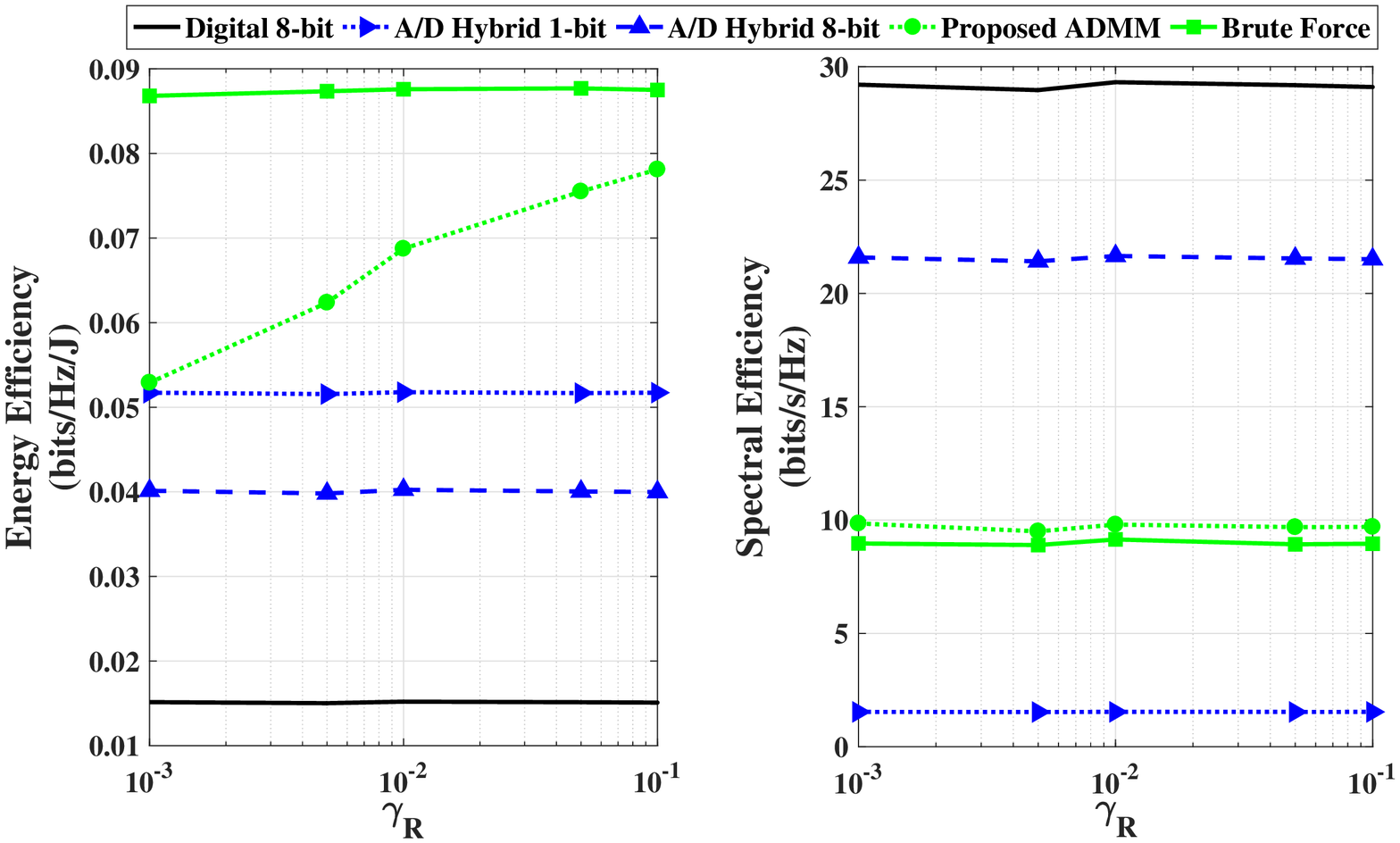}
 		\caption{EE and SE performance w.r.t. $\gamma_\textrm{R}$ at SNR = $10$ dB.}
 \end{figure}
 
 \begin{figure}[t] 
\centering 
\includegraphics[width=0.85\textwidth]{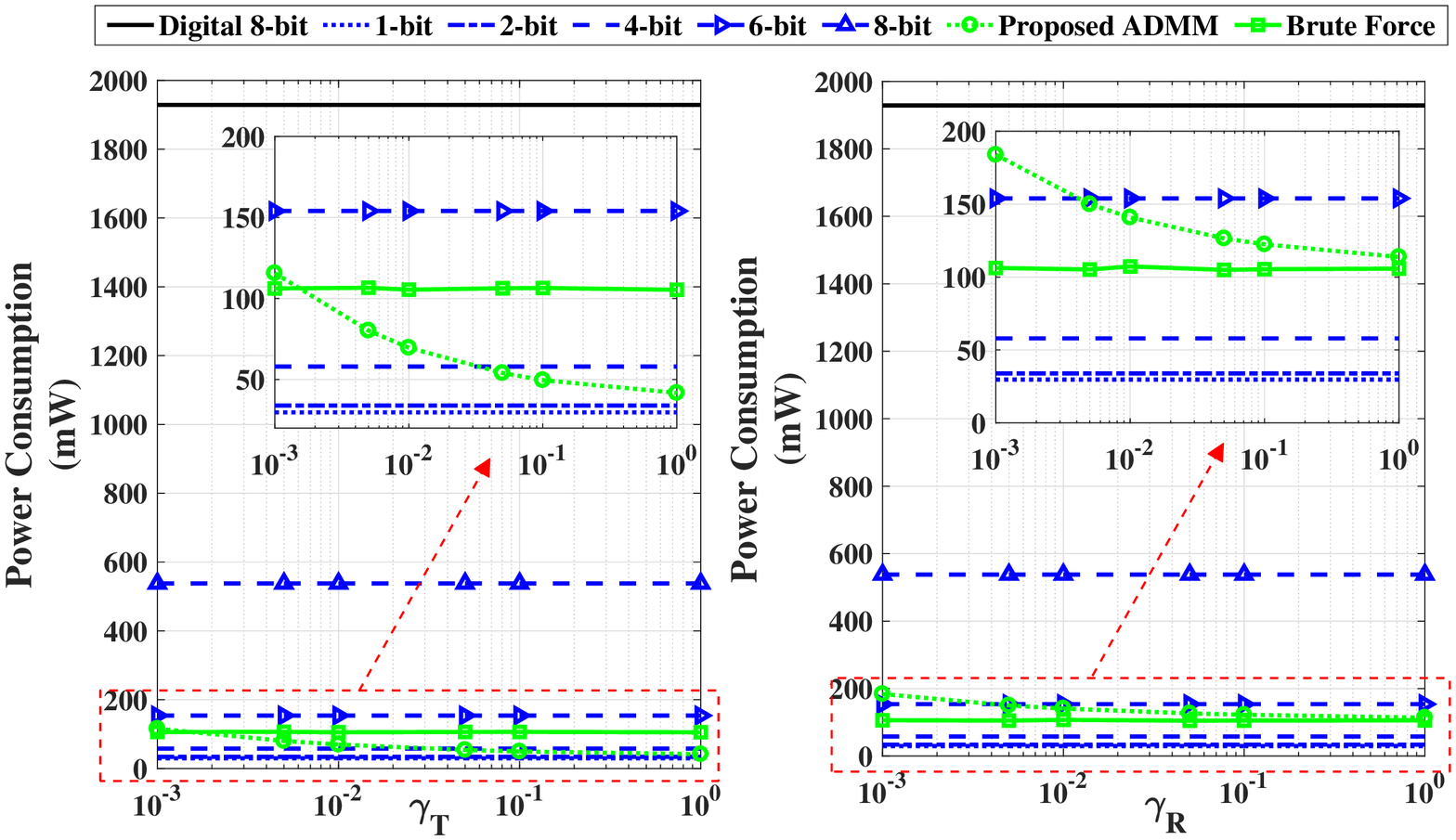}
 		\caption{Power consumption w.r.t. $\gamma_\textrm{T}$ and $\gamma_\textrm{R}$ at the TX and RX, respectively, at SNR = $10$ dB.}
 		\vspace{-4mm}
 \end{figure}

Fig. 3 shows the performance of the proposed ADMM solution compared with existing benchmark techniques w.r.t. SNR at $\gamma_\textrm{T}=0.001$ and $\gamma_\textrm{R}=0.5$. The proposed ADMM solution achieves high EE which is computed by \eqref{eq:ee_problem} after obtaining the optimal DAC and ADC bit resolution matrices, i.e., $\mathbf{\Delta}_\textrm{TX}$ and $\mathbf{\Delta}_\textrm{RX}$, and optimal hybrid precoding and combining matrices, i.e., $\mathbf{F}_\textrm{RF}$, $\mathbf{F}_\textrm{BB}$ and $\mathbf{W}_\textrm{RF}$, $\mathbf{W}_\textrm{BB}$. The results are plugged into \eqref{eq:rate} and \eqref{eq:power_total} to evaluate rate and power  respectively. The EE for the proposed solution has similar performance to the BF approach and is better than the hybrid 1-bit, the hybrid 8-bit and the digital full-bit baselines, e.g., at SNR = $10$ dB, the proposed ADMM solution outperforms the hybrid 1-bit, the hybrid 8-bit and the digital full-bit baselines by about 0.03 bits/Hz/J, 0.04 bits/Hz/J and 0.065 bits/Hz/J, respectively. 

The proposed solution also exhibits better SE, which is the rate in \eqref{eq:rate} after obtaining the optimal DAC and ADC bit resolution matrices, and optimal hybrid precoding and combining matrices, than the hybrid 1-bit and has similar performance to the BF approach for high and low SNR regions and hybrid 8-bit baseline for low SNR region. Note that the proposed ADMM solution enables the selection of different resolutions for different DACs/ADCs and thus, it offers a better trade-off for EE versus SE than existing approaches which are based on a fixed DAC/ADC bit resolution. 
 
Fig. 4 shows the EE (from \eqref{eq:ee_problem}) and SE (from \eqref{eq:rate}) performance results w.r.t. the number of TX antennas $N_\textrm{T}$ at $10$ dB SNR, $\gamma_\textrm{T}=0.001$ and $\gamma_\textrm{R}=0.5$. The proposed ADMM solution again achieves high EE and performs similar to the BF approach and better than the hybrid 1-bit, the hybrid 8-bit and the digital full-bit baselines. For example, at $N_\textrm{T} = 20$, the proposed ADMM solution outperforms hybrid 1-bit, the hybrid 8-bit and the digital full-bit baselines by about 0.03 bits/Hz/J, 0.045 bits/Hz/J and 0.06 bits/Hz/J, respectively. The proposed ADMM solution also exhibits SE performance similar to the BF approach and better than the hybrid 1-bit baseline. 

Fig. 5 shows the EE performance results w.r.t. the number of RX antennas $N_\textrm{R}$ and the number of RX RF chains $L_\textrm{R}$, respectively, at $10$ dB SNR, $\gamma_\textrm{T}=0.001$ and $\gamma_\textrm{R}=0.5$. The proposed ADMM solution again achieves high EE which decreases with increase in the number of RX RF chains, and performs similar to the BF approach (for versus $N_\textrm{R}$) and better than the hybrid 1-bit, the hybrid 8-bit and the digital full-bit baselines. For example, at $N_\textrm{R} = 7$, the proposed ADMM solution outperforms hybrid 1-bit, the hybrid 8-bit and the digital full-bit baselines by about 0.03 bits/Hz/J, 0.06 bits/Hz/J and 0.09 bits/Hz/J, respectively. Also, e.g., at $L_\textrm{R} = 6$, the proposed ADMM solution outperforms hybrid 1-bit, the hybrid 8-bit and the digital full-bit baselines by about 0.025 bits/Hz/J, 0.08 bits/Hz/J and 0.115 bits/Hz/J, respectively. Due to the high complexity of the BF approach, we do not plot results for this approach w.r.t. $L_\textrm{T}$ and $L_\textrm{R}$.

Fig. 6 shows the EE and SE performance results w.r.t. the number of TX RF chains $L_\textrm{T}$ at $10$ dB SNR, $\gamma_\textrm{T}=0.001$ and $\gamma_\textrm{R}=0.5$. The proposed ADMM solution achieves high EE, though this decreases with increase in the number of TX RF chains ADMM achieves better EE performance than the hybrid 1-bit, the hybrid 8-bit and the digital full-bit resolution baselines.  Also, the proposed ADMM solution exhibits SE performance better than the hybrid 1-bit baseline. 

Furthermore, we investigate the performance over the trade-off parameters $\gamma_\textrm{T}$ and $\gamma_\textrm{R}$ introduced in ($\mathcal{P}_\textrm{2}$) and ($\mathcal{P}_\textrm{5}$), respectively. Fig. 7 shows the bar plot of the average of the optimal number of bits selected by the proposed ADMM solution for each DAC versus $\gamma_\textrm{T}$ and for each ADC versus $\gamma_\textrm{R}$. It can be observed that the average optimal number decreases with the increase in $\gamma_\textrm{T}$ and $\gamma_\textrm{R}$, for example, the average number of DAC bits is around 6 for $\gamma_\textrm{T} = 0.001$, 5 for $\gamma_\textrm{T} = 0.01$ and 4 for $\gamma_\textrm{T} = 0.1$. Similarly, at the RX, the average number of ADC bits is about 5 for $\gamma_\textrm{R} = 0.001$, 4 for $\gamma_\textrm{R} = 0.01$ and 3 for $\gamma_\textrm{R} = 0.1$. This is because increasing $\gamma_\textrm{T}$ or $\gamma_\textrm{R}$ gives more weight to the power consumption.

Figs. 8 and 9 show the EE and SE plots for several solutions w.r.t. $\gamma_\textrm{T}$ and $\gamma_\textrm{R}$ at the TX and the RX, respectively. It can be observed that the proposed solution achieves higher EE performance than the fixed bit allocation solutions such as the digital full-bit, the hybrid 1-bit and the hybrid 8-bit baselines and achieves comparable EE and SE results to the BF approach. These curves also show that adjusting $\gamma_\textrm{T}$ and $\gamma_\textrm{R}$ values allow the system to vary the energy-rate trade-off. Note that the TX also accounts for the extra power term, i.e., $\textrm{tr}(\mathbf{F}\mathbf{F}^H)$ as shown in \eqref{eq:tx_power} which means that the selected $\gamma_\textrm{T}$ parameter at the TX is lower than the selected $\gamma_\textrm{R}$ parameter at the RX. Fig. 10 shows that the power consumption in the proposed case is low and decreases with the increase in the trade-off parameter $\gamma_\textrm{T}$ and $\gamma_\textrm{R}$ values unlike digital 8-bit, fixed bit resolution hybrid baselines and the BF approach. 

\section{Conclusion}
This paper proposes an energy efficient mmWave A/D hybrid MIMO system which can vary dynamically the DAC and ADC bit resolutions at the TX and the RX, respectively. This method uses the decomposition of the A/D hybrid precoder/combiner matrix into three parts representing the analog precoder/combiner matrix, the DAC/ADC bit resolution matrix and the digital precoder/combiner matrix. These three matrices are optimized by a novel ADMM solution which outperforms the EE of the digital full-bit, the hybrid 1-bit beamforming and the hybrid 8-bit beamforming baselines, for example, by $3\%$, $4\%$ and $6.5\%$, respectively, for a typical value of $10$ dB SNR. There is an energy-rate trade-off with the BF approach which yields the upper bound for EE maximization and the proposed ADMM solution exhibits lower computational complexity. Moreover, the proposed ADMM solution enables the selection of the optimal resolution for each DAC/ADC and thus, it offers better trade-off for data rate versus EE than existing approaches that are based on a fixed DAC/ADC bit resolution. 
\vspace{-3mm}

\bibliographystyle{IEEEtran}

\begin{thebibliography}{1}
\bibitem{aryanGC2019}
A. Kaushik et al., ``Energy Efficient ADC Bit Allocation and Hybrid Combining for Millimeter Wave MIMO Systems," \emph{IEEE Global Commun. Conf. (GLOBECOM)}, HI, USA, pp. 1-6, Dec. 2019.
\bibitem{andJSAC2014}
J. G. Andrews et al., ``What will 5G be?", \emph{IEEE J. Sel. Areas Commun.}, vol. 32, no. 6, pp. 1065-1082, June 2014.
\bibitem{rappaportACCESS2013}
T. S. Rappaport et al., ``Millimeter wave mobile communications for 5G cellular: It will work!", {\it IEEE Access}, vol. 1, pp. 335-349, 2013.
\bibitem{boccardiCM2014}
F. Boccardi et al., ``Five disruptive technology directions for 5G," {\it IEEE  Commun.  Mag.}, vol. 52, no. 2, pp. 74–80, Feb. 2014.
\bibitem{ayachSPAWC2012}
O. E. Ayach et al., ``The capacity optimality of beam steering in large millimeter wave MIMO systems," {\it IEEE 13th Int. Workshop Signal Process. Advances Wireless Commun. (SPAWC)}, pp. 100-104, June 2012.
\bibitem{ayachTWC2014}
O. E. Ayach et al., ``Spatially sparse precoding in millimeter wave MIMO systems", {\it IEEE Trans. Wireless Commun.}, vol. 13, no. 3, pp. 1499-1513, Mar. 2014.
\bibitem{aryanIET2016}
A. Kaushik et al.,``Sparse hybrid precoding and combining in millimeter wave MIMO systems", in \emph{Proc. IET Radio Prop. Tech. 5G}, Durham, UK, pp. 1-7, Oct. 2016.
\bibitem{hanCM2015}
S. Han et al., ``Large-scale antenna systems with hybrid analog and digital beamforming for millimeter wave 5G," {\it IEEE Commun. Mag.}, vol. 53, no. 1, pp. 186-194, Jan. 2015.
\bibitem{bogaleTWC2016}
T. E. Bogale et al., ``On
the number of RF chains and phase shifters and scheduling design with hybrid analog digital beamforming," {\it IEEE Trans. Wireless Commun.}, vol. 15, no. 5, pp. 3311-3326, May 2016.
\bibitem{payamiTWC2016}
S. Payami et al., ``Hybrid beamforming for large antenna arrays with phase shifter selection," {\it IEEE Trans. Wireless Commun.}, vol. 15, no. 11, pp. 7258-7271, Nov. 2016.
\bibitem{payamiTVT2018}
S. Payami et al., ``Hybrid
beamforming with a reduced number of phase shifters for massive mimo systems," {\it IEEE Trans. Veh. Tech.}, vol. 67, no. 6,
pp. 4843-4851, June 2018.
\bibitem{liCL2017}
A. Li and C. Masouros, ``Hybrid analog-digital millimeter-wave mumimo
transmission with virtual path selection," {\it IEEE Commun. Letters}, vol. 21, no. 2, pp. 438-441, Feb. 2017.
\bibitem{TsinosTSP2018}
C. G. Tsinos et al., ``Hybrid Analog-Digital Transceiver Designs for mmWave Amplify-and-Forward Relaying Systems," {\it Int. Conf. Telecommun. Sig. Process. (TSP), Athens, Greece}, pp. 1-6, 2018.
\bibitem{TsinosASILOMAR2016}
C. G. Tsinos et al., ``Hybrid analog-digital transceiver designs for cognitive radio millimeter wave systems," {\it Asilomar Conf. Sig. Syst. Comput., Pacific Grove, CA}, pp. 1785-1789, 2016.
\bibitem{TsinosTCCN2019}
C. G. Tsinos, S. Chatzinotas and B. Ottersten, ``Hybrid Analog-Digital Transceiver Designs for Multi-User MIMO mmWave Cognitive Radio Systems," {\it IEEE Trans. Cognitive Commun. Netw.}, accepted, Aug. 2019.
\bibitem{ranziSAC2016}
R. Zi et al., ``Energy efficiency optimization of 5G radio frequency chain systems", {\it IEEE J. Sel. Areas Commun.}, vol. 34, no. 4, pp. 758-771, Apr. 2016.
\bibitem{TsinosSAC2017} 
C. G. Tsinos et al., ``On the Energy-Efficiency of Hybrid Analog-Digital Transceivers for Single- and Multi-Carrier Large Antenna Array Systems", \emph{IEEE J. Sel. Areas Commun.}, vol. 35, no. 9, pp. 1980-1995, Sept. 2017.
\bibitem{aryanTGCN2019}
A. Kaushik et al., ``Dynamic RF Chain Selection for Energy Efficient and Low Complexity Hybrid Beamforming in Millimeter Wave MIMO Systems," \emph{IEEE Trans. Green Commun. Netw.}, vol. 3, no. 4, pp. 886-900, Dec. 2019.
\bibitem{aryanComNet2020}
A. Kaushik et al., ``Energy Efficiency Maximization in Millimeter Wave
Hybrid MIMO Systems for 5G and Beyond," \emph{IEEE Int. Conf. Commun. Netw. (ComNet)}, pp. 1-7, Mar. 2020.
\bibitem{heathSSP2016}
R. W. Heath et al., ``An overview of signal processing techniques for millimeter wave MIMO systems", \emph{IEEE J. Sel. Topics Signal Process.}, vol. 10, no. 3, pp. 436-453, Apr. 2016.
\bibitem{orhanITA2015}
O. Orhan et al., ``Low power analog-to-digital conversion in millimeter wave systems: Impact of resolution and bandwidth on performance", \emph{2015 Info. Theory Appl. Workshop, San Diego, CA}, pp. 191-198, 2015.
\bibitem{fanCL2015}
L. Fan et al., ``Uplink achievable rate for massive MIMO systems with low-resolution ADC", {\it IEEE Commun. Letters}, vol. 19, no. 12, pp. 2186-2189, Oct. 2015.
\bibitem{choiTSP2017}
J. Choi et al., ``Resolution-adaptive hybrid MIMO architectures
for millimeter wave communications", \emph{IEEE Trans. Sig. Process.}, vol. 65, no. 23, pp. 6201-6216, Dec. 2017.
\bibitem{jmoITG2016}
J. Mo et al., ``Achievable rates of hybrid architectures with few-bit ADC receivers," \emph{VDE Int. ITG Workshop Smart Antennas}, pp. 1-8, 2016.
\bibitem{zhangSAC2017}
J. Zhang et al., ``Performance analysis of mixed-ADC massive MIMO systems over Rician fading channels," \emph{IEEE J. Sel. Areas Commun.}, vol. 35, no. 6, pp. 1327-1338, Jun. 2017.
\bibitem{aryanEUSIPCO2018}
A. Kaushik et al.,``Efficient channel estimation in millimeter wave hybrid MIMO systems with low resolution ADCs", \emph{IEEE Europ. Sig. Process.}, Rome, Italy, pp. 1839-1843, Sept. 2018.
\bibitem{tczhangTWC2016}
T.-C. Zhang et al., ``Mixed-ADC massive MIMO detectors: Performance analysis and design optimization," \emph{IEEE Trans. Wireless Commun.}, vol. 15, no. 11, pp. 7738-7752, Nov. 2016.
\bibitem{aryanICC2019}
A. Kaushik et al., ``Energy Efficiency maximization of millimeter wave hybrid MIMO systems with low resolution DACs," \emph{IEEE Int. Conf. Commun. (ICC)}, Shanghai, China, pp. 1-6, May 2019.
\bibitem{singhTWC2009}
J. Singh, et al., ``On the limits of communication
with low-precision analog-to-digital conversion at the receiver,” \emph{IEEE Trans. Wireless Commun.}, vol. 57, no. 12, pp. 3629-3639, Dec. 2009.
\bibitem{mezghaniECS2009}
A. Mezghani et al., ``Transmit processing with low resolution D/A-converters", \emph{Int. Conf. Electronics, Circuits  Systems}, Tunisia, pp. 683-686, Dec. 2009.
\bibitem{jacobssonTC2017}
S. Jacobsson et al., ``Quantized precoding for massive MU-MIMO", in \emph{IEEE Trans. Commun.},  vol. 65, no. 11, pp. 4670-4684, Nov. 2017.
\bibitem{TsinosSPAWC2018}
C. G. Tsinos et al., ``Symbol-Level Precoding with Low Resolution DACs for Large-Scale Array MU-MIMO Systems," {\it Int. Workshop Sig. Process. Adv. Wireless Commun., Kalamata, Greece}, pp. 1-5, 2018.
\bibitem{ribeiroTSP2018}
L. N. Ribeiro et al., "Energy efficiency of mmWave massive MIMO precoding with low-resolution DACs," in \emph{IEEE J. Sel. Topics Sig. Process.}, vol. 12, no. 2, pp. 298-312, May 2018.
\bibitem{ADMM}
S. Boyd et al. ``Distributed optimization and statistical learning via the alternating direction method of multipliers," \emph{Foundations and Trends in Machine Learning}, vol. 3, no. 1, pp. 1-122, 2011.
\bibitem{singh}
S. Singh et al., ``Interference analysis for highly directional 60-GHz mesh networks: The case for rethinking medium access control", \textit{IEEE/ACM Trans. Netw.}, vol. 19, no. 5, pp. 1513-1527, Oct. 2011.
\bibitem{brady}
J. Brady et al., ``Beamspace MIMO for millimeter-wave communications: system architecture, modeling, analysis, and measurements", {\it IEEE Trans. Antenn. Propag.}, vol. 61, no. 7, pp. 3814-3827, Jul. 2013.
\bibitem{dai}
L. Dai et al., ``Beamspace channel estimation for millimeter-wave massive MIMO systems with lens antenna array", in {\it 2016 IEEE/CIC Int. Conf. Commun. China (ICCC)}, pp. 1-6, July 2016.
\bibitem{mezghaniISIT2012}
A. Mezghani and J. A. Nossek, ``Capacity lower bound of MIMO channels with output quantization and correlated noise," \emph{IEEE Int. Symp. Info. Theory (ISIT)}, Cambridge, USA, Jul. 2012.
\bibitem{zapponeFTCIT2015}
A. Zappone and E. Jorswieck, ``Energy Efficiency in Wireless Networks via Fractional Programming Theory," \emph{Found. Trends Commun. Info. Theory}, vol. 11, no. 3-4, pp 185-396, 2015. 
\bibitem{palomarJSAC2006}
D. P. Palomar and M. Chiang, ``A tutorial on decomposition methods for network utility maximization," \emph{IEEE J. Sel. Areas Commun.}, vol. 24,no. 8, pp. 1439-1451, Aug. 2006.
\bibitem{dinkelbach}
W. Dinkelbach, ``On nonlinear fractional programming", {\it Management Science}, vol. 13, no. 7, pp. 492-498, Mar. 1967.
\bibitem{tse2004}
D. Tse and P. Viswanath, Fundamentals of Wireless Communication, Cambridge University Press, UK, 2004.
\bibitem{TSINOSNC1}
C. G. Tsinos et al., ``Distributed blind hyperspectral unmixing via joint sparsity and
low-rank constrained non-negative matrix factorization,'' \emph{IEEE Trans. Comput. Imag.}, vol. 3, no. 2, pp. 160-174, June
2017.
\bibitem{TSINOSNC2}
C. G. Tsinos and B. Ottersten, ``An efficient algorithm for unit-modulus quadratic programs with application in
beamforming for wireless sensor networks,'' \emph{IEEE Sig. Process. Letters}, vol. 25, no. 2, pp. 169-173, Feb. 2018.
\bibitem{AGL}
D. P. Bertsekas, ``Nonlinear programming," Athena Scientific, USA, Sept. 1999.
\bibitem{BER}
J. Eckstein and D. P. Bertsekas, ``On the DouglasRachford splitting method and the proximal point algorithm for maximal monotone
operators," \emph{Mathematical Programming}, vol. 55, no. 1-3, pp. 293-318, 1992.
\bibitem{GOL}
E. G. Golshtein and N. Tretyakov, ``Modified lagrangians in convex programming and their generalizations," in \emph{Point-to-Set Maps and Mathematical Programming, Springer}, pp. 86-97, 1979.
\bibitem{cvx}
M. Grant and S. Boyd, ``Graph implementations for nonsmooth convex programs", in \emph{Recent Adv. Learning and Control}, Springer-Verlag Ltd., pp. 95-110, 2008.
\bibitem{rappa2}
T. S. Rappaport et al., ``Millimeter wave wireless communications," Prentice-Hall, NJ, USA, Sept. 2014.
\end{thebibliography}

%
\vskip -2\baselineskip plus -1fil
\begin{IEEEbiography}[{\includegraphics[width=1in,height=1.25in,clip,keepaspectratio]{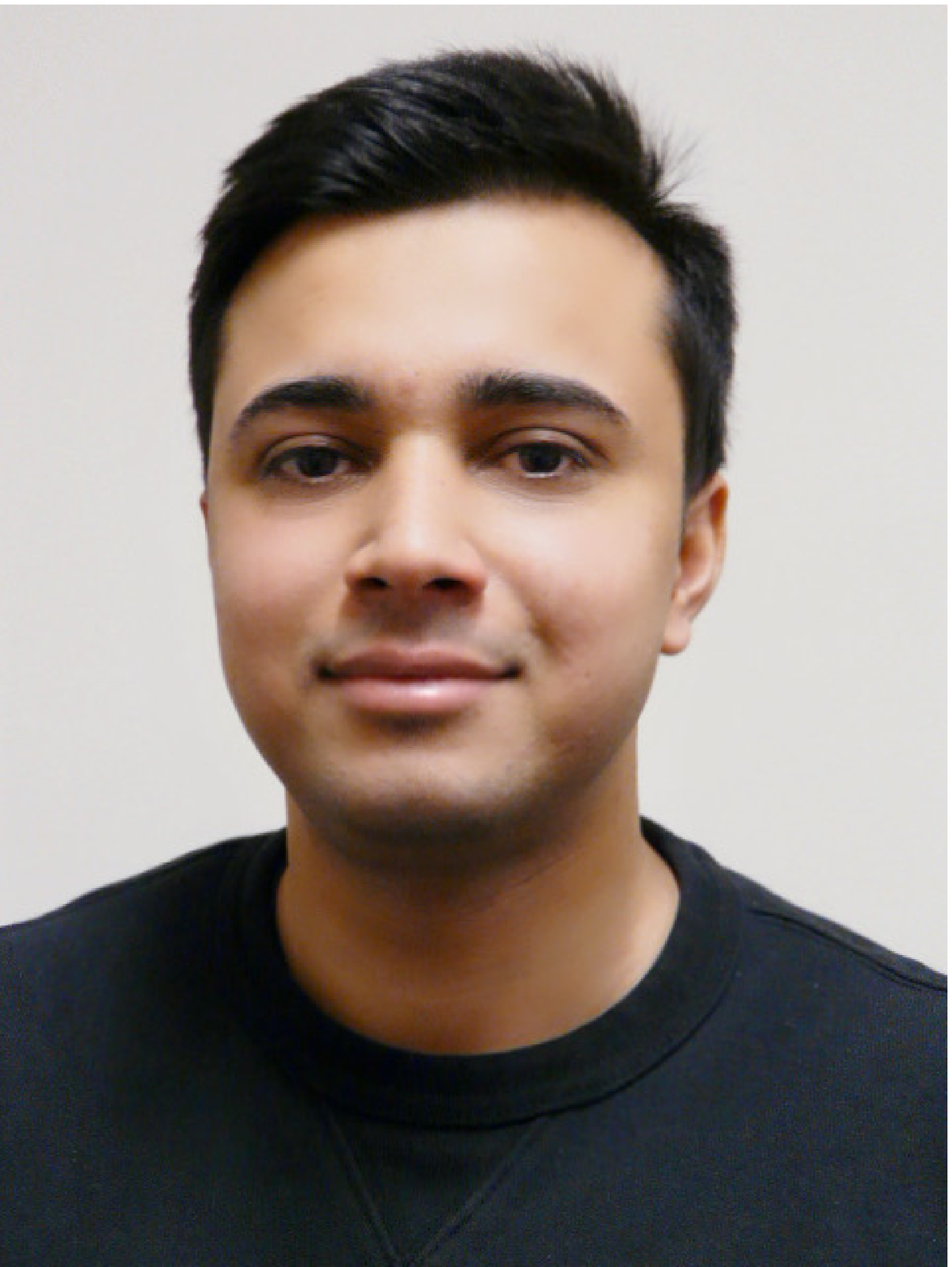}}]{Aryan Kaushik} is a Research Fellow in Communications and Radar Transmission at the Department of Electronic and Electrical Engineering, University College London, U.K. He received his Ph.D. degree in communications engineering at the Institute for Digital Communications, The University of Edinburgh, U.K., in 2020.
He received his M.Sc. degree in telecommunications from The Hong Kong University of Science and Technology, Hong Kong, in 2015. He has held visiting research appointments at the Imperial College London, U.K., from 2019-20, University of Luxembourg, Luxembourg, in 2018, and Beihang University, China, in the period of 2017-19. 
His research interests include signal processing for communications, dual communications and radar transmission, energy efficient wireless communications and millimeter wave massive MIMO systems.
\end{IEEEbiography}
\vskip -2\baselineskip plus -1fil
\begin{IEEEbiography}[{\includegraphics[width=1in,height=1.25in,clip,keepaspectratio]{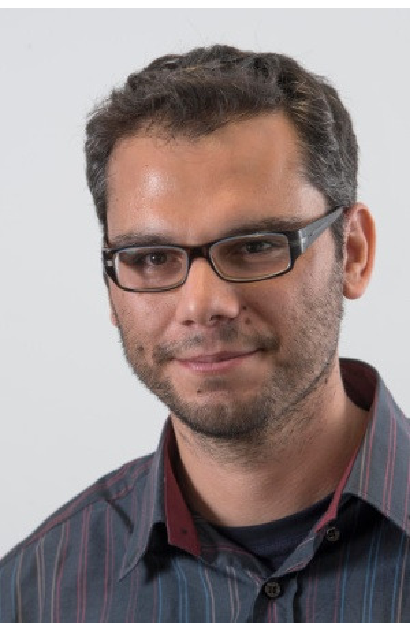}}]{Evangelos Vlachos (M'19)} is a Research Associate at the Industrial Systems Institute, Athena Research Centre, Patras, Greece. He has been a Research Associate in Signal Processing for Communications at the Institute for Digital Communications, The University of Edinburgh, U.K., from 2017-19. He received the Diploma,
M.Sc. and Ph.D. degrees from the Computer Engineering and Informatics Department, 
University
of Patras, Greece, in 2005, 2009, and 2015, respectively. 
From 2015-16, he was a Postdoctoral Researcher at the Laboratory of Signal Processing and Telecommunications in the University of Patras, Greece, 
where in 2016, he worked 
at the Visualization and Virtual Reality Group. 
He received best paper award from IEEE ICME in 2017 and participated in six research projects funded by the EU. 
His research interests include wireless communications, machine learning and optimization, adaptive control and filtering algorithms.
\end{IEEEbiography}
\vskip -2\baselineskip plus -1fil
\begin{IEEEbiography}[{\includegraphics[width=1in,height=1.25in,clip,keepaspectratio]{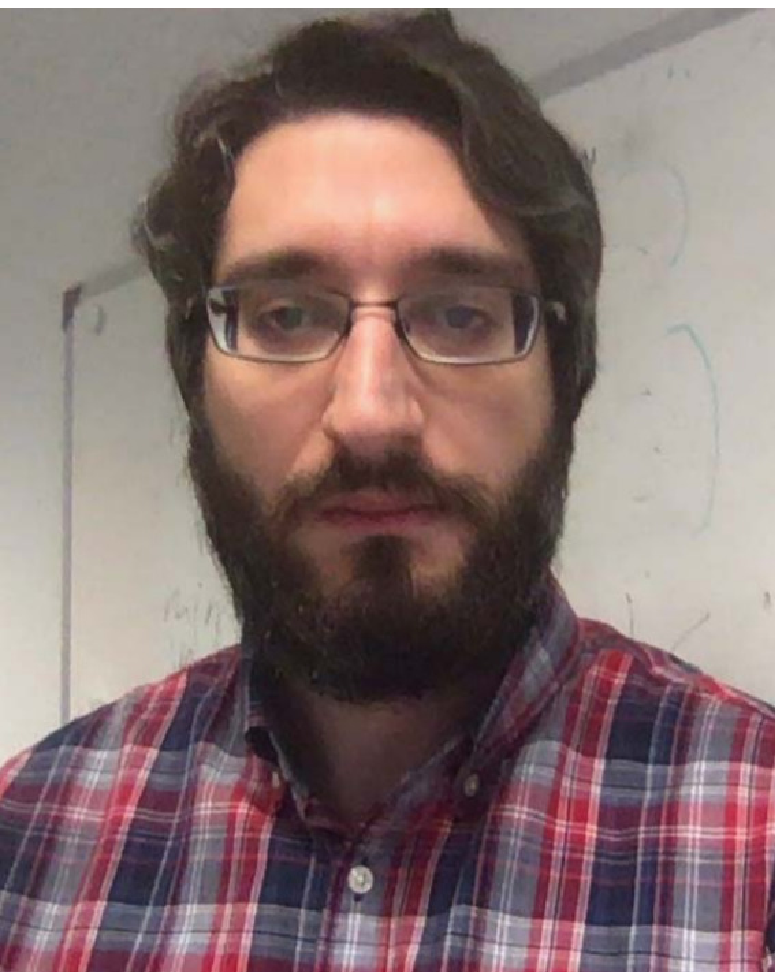}}]{Christos Tsinos (S'08-M'14)} is a Research Associate at the Interdisciplinary Centre for Security, Reliability and Trust, University of Luxembourg, Luxembourg. He received the
Diploma degree in computer engineering and informatics, M.Sc. and Ph.D. degrees in signal processing and communication systems, and M.Sc.
degree in applied mathematics from the University
of Patras, Greece, in 2006, 2008, 2013, and 2014,
respectively. From 2014-15, he was a Postdoctoral Researcher at the University of Patras, Greece. He is currently the Principal Investigator of the project ``Energy and CompLexity EffiCienT mIllimeter-wave large-array Communications (ECLECTIC)" funded under FNR CORE Framework and a member of the Technical Chamber of Greece. His research interests include signal processing for millimeter wave, massive MIMO, cognitive radio, cooperative and satellite communications, wireless sensor networks and hyperspectral image processing.
\end{IEEEbiography}
\vskip -2\baselineskip plus -1fil
\begin{IEEEbiography}[{\includegraphics[width=1in,height=1.25in,clip,keepaspectratio]{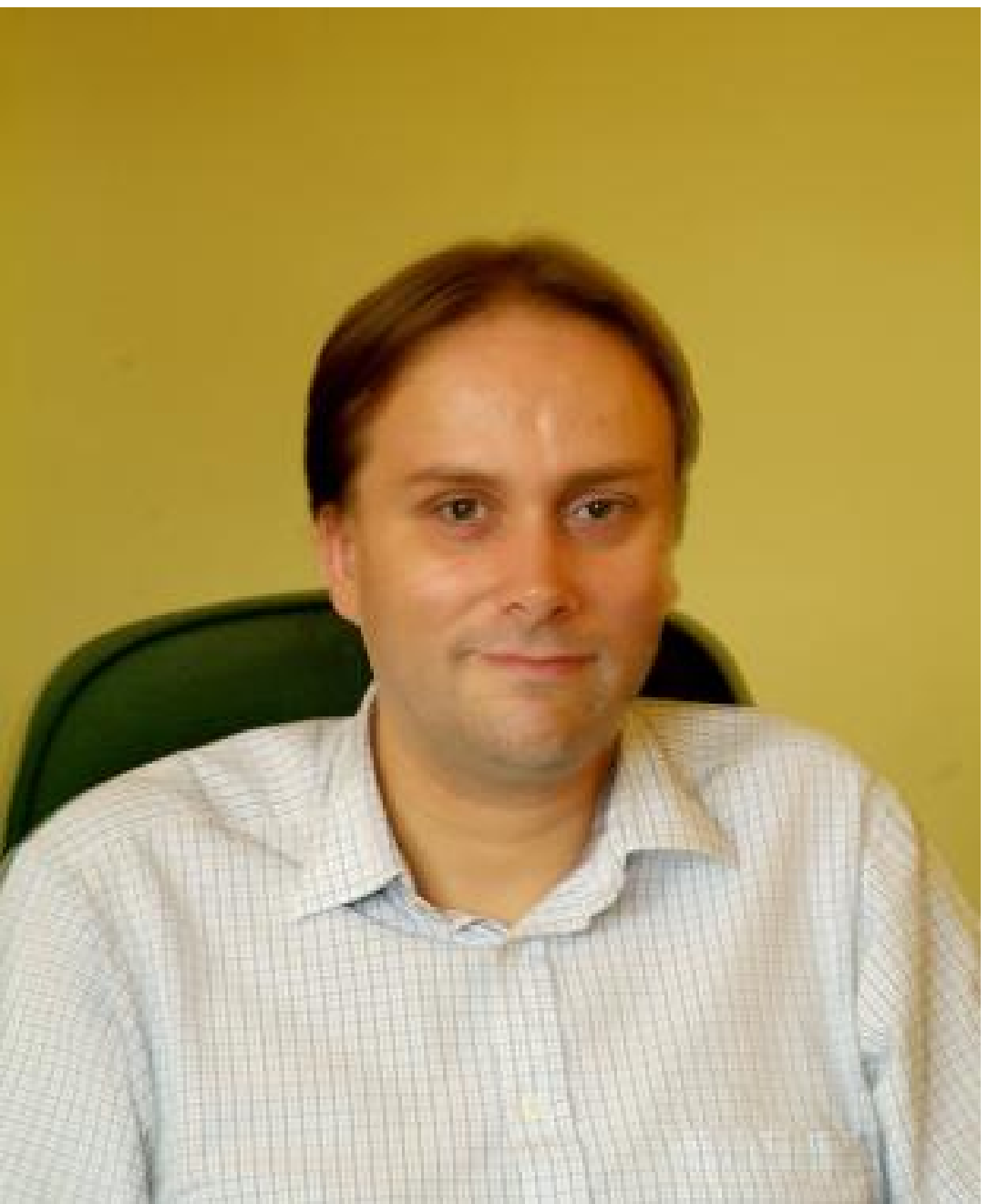}}]{John Thompson (S'94-M'03-SM'13-F'16)} is a Professor of Signal Processing and Communications and Director of Discipline at The University of Edinburgh, U.K. He is listed by Thomson Reuters as a highly cited scientist from 2015-18. He specializes in millimetre wave wireless communications, signal processing for wireless networks, smart grid concepts for energy efficiency green communications systems and networks, and rapid prototyping of MIMO detection algorithms. He has published over 300 journal and conference papers on these topics. He co-authored the second edition of the book entitled ``Digital Signal Processing: Concepts and Applications''. He coordinated EU Marie Curie International Training Network ADVANTAGE on smart grid from 2014-17. He is an Editor for IEEE Transactions on Green Communications and Networking, and Communications Magazine Green Series, Former founding Editor-in-Chief of IET Signal Processing, Technical Programme Co-chair for IEEE Communication Society ICC 2007 Conference and Globecom 2010 Conference, Technical Programme Co-chair for IEEE Vehicular Technology Society VTC Spring 2013 Conference, Track co-chair for the Selected Areas in Communications Topic on Green Communication Systems and Networks at ICC 2014 Conference, Member at Large of IEEE Communications Society Board of Governors from 2012-2014, Tutorial co-chair for IEEE ICC 2015 Conference, Technical programme co-chair for IEEE Smartgridcomm 2018 Conference. Tutorial co-chair for ICC 2015 Conference. He is the local student counsellor for the IET and local liaison officer for the UK Communications Chapter of the IEEE.
\end{IEEEbiography}
\vskip -2\baselineskip plus -1fil
\begin{IEEEbiography}[{\includegraphics[width=1in,height=1.25in,clip,keepaspectratio]{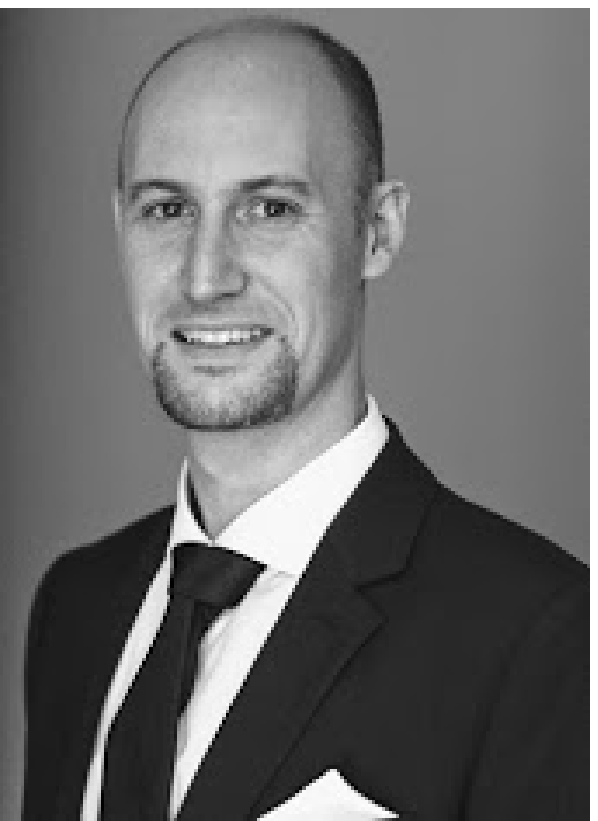}}]{Symeon Chatzinotas (S'06-M'09-SM'13)} is a Professor and Co-Head of the SIGCOM group in the Interdisciplinary Centre for Security, Reliability and Trust, University of Luxembourg, Luxembourg. He has been also a Visiting Professor at the University of Parma, Italy. His research interests are on multiuser information theory, cooperative/cognitive communications, cross-layer wireless network optimization and content delivery networks. In the past, he has worked in numerous Research and Development projects for the Institute of Informatics and Telecommunications, National Center for Scientific Research ``Demokritos", the Institute of Telematics and Informatics, Center of Research and Technology Hellas, and Mobile Communications Research Group, Center of Communication Systems Research, University of Surrey, U.K. He has authored more than 300 technical papers in refereed international journals, conferences and scientific books. He was a co-recipient of the 2014 IEEE Distinguished Contributions
to Satellite Communications Award, the CROWNCOM 2015 Best Paper
Award, 2018 EURASIP JWCN Best Paper Award and 2019 ICSSC Best Student Paper Award.
\end{IEEEbiography}
\end{document}